\documentclass[sigconf]{acmart}

\AtBeginDocument{%
  }

\copyrightyear{2023}
\acmYear{2023}
\setcopyright{acmlicensed}
\acmConference[KDD '23] {Proceedings of the 29th ACM SIGKDD Conference on Knowledge Discovery and Data Mining}{August 6--10, 2023}{Long Beach, CA, USA.}
\acmBooktitle{Proceedings of the 29th ACM SIGKDD Conference on Knowledge Discovery and Data Mining (KDD '23), August 6--10, 2023, Long Beach, CA, USA}
\acmPrice{15.00}
\acmISBN{979-8-4007-0103-0/23/08}
\acmDOI{10.1145/3580305.3599540}

\usepackage{graphicx}
\usepackage{url}
\usepackage{comment}
\usepackage{xspace}
\usepackage{amsmath}
\usepackage{xcolor}

\usepackage[noend]{algpseudocode}

\usepackage{algorithm}
\usepackage{multirow}
\usepackage{soul}
\usepackage{subcaption}
\usepackage[noabbrev, capitalise]{cleveref}

\theoremstyle{acm}
\newtheorem{definition}{Definition}[section]

\begin{document}

\title{Using Motif Transitions for Temporal Graph Generation}
\author{Penghang Liu}
\affiliation{%
  \institution{University at Buffalo}
  \city{Buffalo}
  \state{NY}
  \country{USA}}
\email{penghang@buffalo.edu}
\orcid{0000-0002-1238-7738}

\author{Ahmet Erdem Sar{\i}y{\"u}ce}
\affiliation{%
  \institution{University at Buffalo}
  \city{Buffalo}
  \state{NY}
  \country{USA}}
\email{erdem@buffalo.edu}
\orcid{0000-0002-4945-6821}

\begin{abstract}
Graph generative models are highly important for sharing surrogate data and benchmarking purposes.
Real-world complex systems often exhibit dynamic nature, where the interactions among nodes change over time in the form of a temporal network.
Most temporal network generation models extend the static graph generation models by incorporating temporality in the generation process.
More recently, temporal motifs are used to generate temporal networks with better success.
However, existing models are often restricted to a small set of predefined motif patterns due to the high computational cost of counting temporal motifs. 
In this work, we develop a practical temporal graph generator, Motif Transition Model (\textit{MTM}), to generate synthetic temporal networks with realistic global and local features.
Our key idea is modeling the arrival of new events as temporal motif transition processes.
We first calculate the transition properties from the input graph and then simulate the motif transition processes based on the transition probabilities and transition rates.
We demonstrate that our model consistently outperforms the baselines with respect to preserving various global and local temporal graph statistics and runtime performance.

\end{abstract}

\ccsdesc[500]{Theory of computation~Dynamic graph algorithms}
\ccsdesc[500]{Computing methodologies~Network science}

\keywords{temporal networks, graph generative model, temporal motifs}

\maketitle

\section{Introduction}
Graphs are an effective tool to model real-world complex systems in various domains, such as communications, human interactions, financial activities, social relations, and protein interactions.
One important problem is the generation of realistic synthetic networks from a given distribution of real-world networks~\cite{erdHos1959random,chung2002average,chakrabarti2006graph}.
Graph generators are necessary for enabling the sharing of surrogate data, such as computer network traffic or financial networks, and benchmarking studies, such as scalability and versatility tests~\cite{BTER14}.

Real-world complex systems often exhibit dynamic nature, where the interactions among nodes evolve over time.
Edges are active only at certain points in time in temporal networks, such as financial transactions, communication networks, and face-to-face contacts.
Temporal graph generation is relatively understudied and particularly more challenging than static graph generation due to the diverse and large nature of real-world temporal networks.
Most temporal network generation models focus on reproducing the global characteristics of real-world networks.
They often extend the static graph generation models by incorporating temporality in the generation process~\cite{karrer2009random,hill2010dynamic,holme2012temporal,holme2013epidemiologically}. 
Existing models lack a holistic approach that can consider the structural and temporal characteristics at the same time.

\begin{figure}[t!]
\centering
\includegraphics[width=0.95\linewidth]{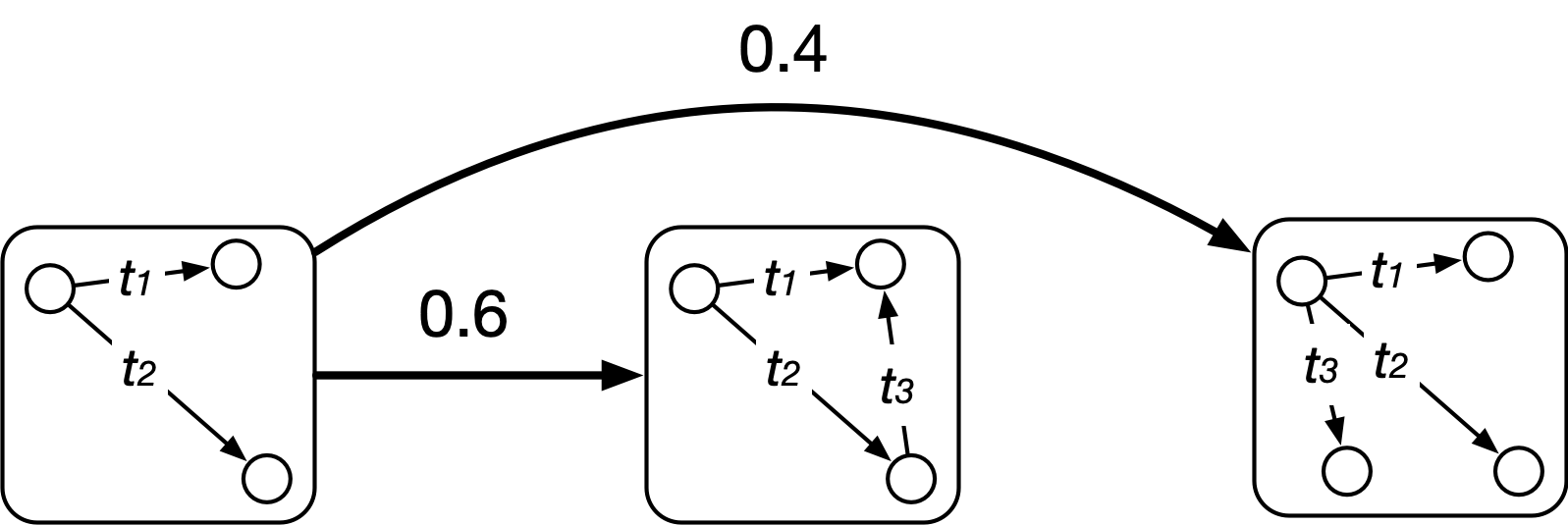}
\vspace{-1ex}
\caption{\small An example of motif transition. A wedge motif can have 0.6 probability of evolving into a triangle and 0.4 probability of becoming a 3-star.}
\vspace{-5ex}
\label{fig:toy}
\end{figure}

Temporal motifs, also known as higher-order structures, are an important building block in temporal networks~\cite{K11, S14, H15, P17, liu2021temporal, porter2022analytical}.
Temporal motif-based analysis has been used for many applications including cattle trade movements~\cite{bajardi2011dynamical}, editor interactions in Wikipedia~\cite{jurgens2012temporal}, mobile communication networks~\cite{kovanen2013temporal, li2014statistically}, and human interactions~\cite{zhang2015human}.
Temporal motifs are also a promising and effective tool for temporal graph generation.
Purohit et al. introduced the Structural Temporal Modeling ({\it STM}) which computes the frequencies of a set of atomic motifs and places the motifs with the preferential attachment mechanism to generate temporal graphs~\cite{purohit2018temporal}.
Zeno et al. \cite{zeno2021dymond} modeled the active node behavior over time and generate three types of 3-node motifs (one edge, wedge, and triangle) with different arrival rates.
However, these models suffer from the cost of counting temporal motifs which increases exponentially as the size of the motif increases. Thus, they are often limited to a custom set of motifs of limited size which are unable to capture the structural and temporal characteristics of the networks.

In this work, we develop a realistic and practical temporal graph generator.
For a given input temporal network, our Motif Transition Model (\textit{MTM}) generates a synthetic temporal network that preserves the global and local features of the input network.
We introduce the motif transition to model the evolution of temporal motifs as a stochastic process.
\cref{fig:toy} gives an example where a wedge motif can transition into two larger temporal motifs by adding new events.
Our model first calculates the transition properties from the input graph and then simulates the stochastic motif transition processes based on the transition probabilities and transition rates.
We compare the \textit{MTM} against several state-of-the-art temporal graph generative models on real-world networks from various domains.
We demonstrate that our model consistently outperforms the baselines with respect to preserving various global and local temporal graph statistics.
Furthermore, our model is significantly more efficient than the existing baselines.

Our key contributions can be summarized as follows:
\begin{itemize}
\item We propose the Motif Transition Model, \textit{MTM}, to generate a synthetic network through a stochastic process that accurately captures the global graph statistics and the temporal motif structures.
\item We give algorithms to efficiently compute the motif transition properties and use those to generate the new graph. In particular, we do not count the motifs in the process which makes our model practical and scalable.
\item We perform a comprehensive evaluation of our model against several baselines on various real-world networks. {\it MTM} outperforms the baselines in preserving the global graph characteristics and local temporal motif features of the input graph. {\it MTM} is also highly efficient and scalable to large networks.
\end{itemize}

\section{Related Work}
Here we briefly summarize the previous work on the generative models for temporal networks and temporal network motifs.

\vspace{-2ex}
\subsection{Generative Models for Temporal Networks}
Most of the previous studies on generative models for temporal networks are derived from the extension of static graph generation models \cite{hill2010dynamic, holme2012temporal, holme2013epidemiologically}. 
Gauvin et al. \cite{gauvin2018randomized} surveyed several random reference models for temporal networks many of which extend the Erd\H{o}s-Renyi and configuration models to temporal networks.
One of the most commonly used static graph generative models is the stochastic block model \cite{airoldi2008mixed}, which splits nodes into different groups and then places edges between them according to their categories. Several dynamic variants of the stochastic block model are developed \cite{xing2010state,ho2011evolving,yang2011detecting,kim2013nonparametric,xu2013dynamic,ghasemian2016detectability,matias2017statistical,zhang2017random}, in which the nodes can switch classes over time. 
Recently, Porter et al. \cite{porter2022analytical} developed the temporal activity state block model \textit{TASBM} to generate a temporal network with temporal motif structures. They divide the temporal graph into multiple time windows and compute the average in-event and out-event arrival rates for each node. Based on the event arrival rates between the activity groups, the model generates the events between each pair of nodes by a Poisson draw.

Inspired by the success of deep graph generative models, Zhou et al. \cite{zhou2020data} proposed the \textit{TagGen} model to generate temporal graphs. They use a bi-level self-attention mechanism and assemble the temporal random walk sequences to form a temporal interaction network. However, their model converts the temporal graph into a few snapshots, hence the fine-grained timestamp information is lost during this process.
The prohibitive time and space complexity also make their model impractical for large scale.
Another line of work used higher-order and variable-order Markov chains to model the temporal sequence of edges in multiple k-th order pathways \cite{peixoto2017modelling,scholtes2017network}.
Most existing temporal graph generative models do not consider the higher-order structures except temporal walks and pathways. However, many real-world events do not occur in traversal patterns, such as repetitive monthly auto-payment activities, which cannot be reproduced by the random walks or higher-order pathways.
In our work, we propose an effective solution to model any potential correlation between two events that are topologically and temporally close to each other.
Our model is also practical for large-scale networks.

\vspace{-2ex}
\subsection{Temporal Motifs}
Higher-order temporal subgraph structures, i.e., temporal motifs, are an important property of temporal networks.
Several approaches have been proposed to model temporal motifs \cite{K11, S14, H15, P17, liu2021temporal}.
Applications and use cases for temporal motifs are numerous.
Jin et al. \cite{jin2007trend} proposed trend motifs to investigate the financial and protein networks with dynamic node weights.
Zhao et al. \cite{zhao2010communication} devised communication motifs to study the information propagation in human communication networks, such as call detail records (CDR) and Facebook wall post interactions.
Zhang et al. \cite{zhang2015human} introduced motif-driven analysis for human interactions including phone messages, face-to-face interactions, and sexual contacts.
Liu et al. studied patent oppositions and collaborations~\cite{Liu22} and financial transaction networks~\cite{Liu23} by using temporal motifs.

Regarding the temporal graph generative models, Purohit et al. used temporal motifs to propose Structural Temporal Modeling (\textit{STM}) process to generate temporal networks~\cite{purohit2018temporal}. They select a set of easy-to-compute atomic motifs, such as wedges, triangles, and squares, and for each type of atomic motif, the model calculates the independent motif frequencies (ITeM) which prohibits the overlaps between motifs~\cite{purohit2022item}.
Nodes and motifs in the output graph are generated based on the ITeM frequency and a preferential attachment function.
Similarly, Zeno et al. \cite{zeno2021dymond} proposed the DYMOND model to generate dynamic networks. They consider three types of undirected static motifs (one edge, wedge, and triangle), and place motifs on active nodes with different arrival rates. However, they convert the temporal network into a sequence of snapshots and only model static motifs which do not contain any timestamp information.
One limitation of these models is that they only select a limited set of temporal motifs (mostly wedges and triangles), which are not sufficient to capture more complex subgraph structures. Another drawback is that they rely on counting the number of temporal motifs and do not consider the correlations between temporal motifs of different sizes. 
Also, the complexity of counting temporal motifs increases exponentially as the motif size increases, thus most existing studies does not consider temporal motifs with more than three events.
In our work, we use the full spectrum of temporal motifs and utilize the transitions among them to generate realistic synthetic temporal networks---we also do not count the motifs in the process which makes our model practical and scalable.

\vspace{-1ex}
\section{Background}
Here we provide the formal definition of temporal motifs, and introduce a notation method to denote different types of temporal motifs. The notations and symbols are summarized in \cref{tab:notation} of Appendix A.

\vspace{-1ex}
\subsection{Temporal Motifs}
We explore temporal networks which we represent as $G = (V, E)$. $V$ is the set of nodes and $E$ is the set of timestamped events.
Each event $e_i \in E$ is a 3-tuple $(u_i, v_i, t_i)$ that represents a directed relation from the source node $u_i$ to the target node $v_i$ at time $t_i$.
The event set $E$ is a time-ordered list of events such that $t_1 < t_2 < t_3 < \cdots < t_{|E|}$.
Here we distinguish edges and events, where the edge $(u, v)$ is the static projection of an event $(u, v, t)$, and there may be multiple events occurring on the same edge at different times. We also define the static projection of $G$ as $\overline{G} = \{(u_i, v_i)\ | \forall (u_i, v_i, t_i) \in G\}$.
\begin{definition}\label{def:motif}{\bf (Temporal motif)} 
Given a temporal network $G = (V, E)$, an $l$-event temporal motif ($l \geq 2$), denoted by $M^l_i = (V', E')$, is a temporal subgraph in $G$ such that
\begin{itemize}
\item $V' \subseteq V$, $E' \subseteq E$, $|E'| = l$,
\item $\overline{M^l}$ is a (weakly-connected) subgraph, hence $2 \le |V'| \le l+1$,
\item Each event is connected to at least one of the previous events: $\{u'_{j+1}, v'_{j+1}\} \cap \{u'_1, v'_1, \dots, u'_j, v'_j\} \neq \emptyset$ for any $j+1 \le l$ (i.e., the motif is a connected subgraph at every timestamp).
\end{itemize}
\end{definition}

For a given $l$, there are different types of temporal motifs in terms of connectivity and temporal structure, called as the motif spectrum.
For instance, there are 6 types in the motif spectrum for $M^2$ (see~\cref{fig:notation}), 60 types for $M^3$ (see~\cref{fig:m60} in Appendix C), and 888 types for $M^4$.
We denote each type of temporal motif with a unique subscript such as $M^2_i$ for $1\le i \le 6$, $M^3_i$ for $1\le i \le 60$, and $M^4_i$ for $1\le i \le 888$.
We use $\{M^l_i\}$ to denote the set of instances of a motif $M^l_i$ in $G$.

\begin{figure}[b!]
\vspace{-3ex}
\centering
\includegraphics[width=0.8\linewidth]{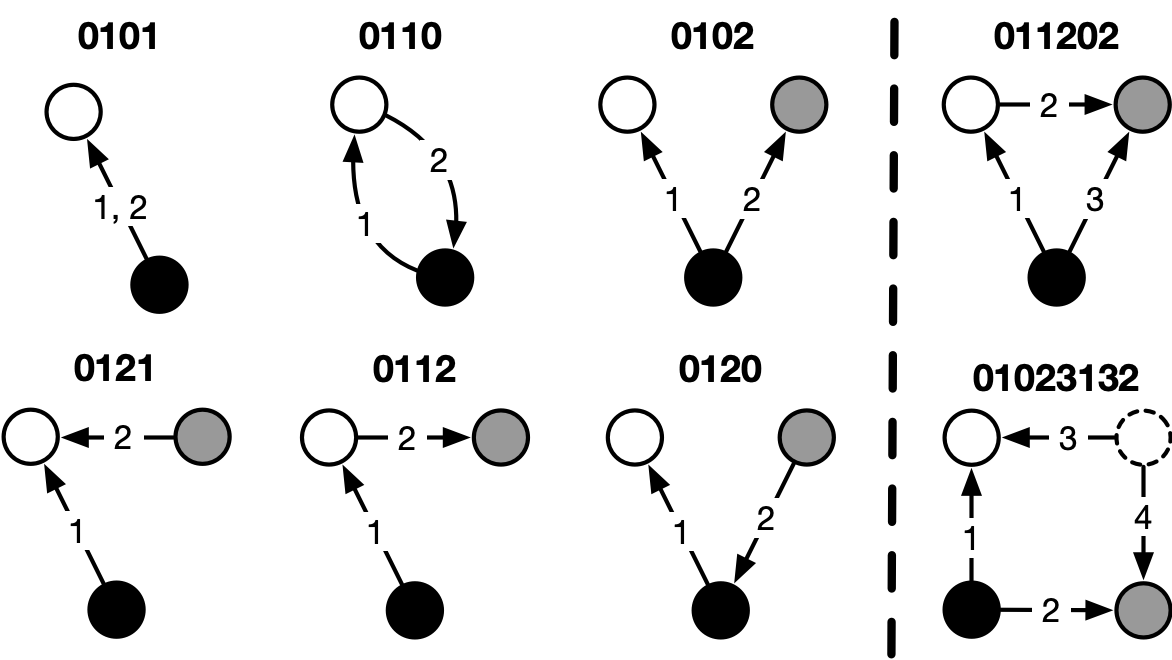}
\vspace{-1ex}
\caption{\small Six types of 2-event motifs (left) and an example of 3-event and 4-event motifs (right). We use $2l$ digits to denote a temporal motif with $l$ events.
Each event is given by a pair of digits, where the source node is the first and the target node is the second digit.
}
\label{fig:notation}
\end{figure}

\subsection{Motif Notation}
To describe the large spectrum of temporal motifs (e.g., $M^3_i$ for $1 \le i \le 60$), we introduce a digit-based notation.
We use $2l$ digits to denote a unique type of $l$-event temporal motif $M^l_i$.
Each pair of digits, $uv$, denotes an event from the node represented by the first digit $u$ to the node denoted by the second digit $v$.
Digits start from zero and the digit for each node follows the chronological order of the node's appearance in the motif.
The sequence of digit pairs also follows the chronological order of the events.
The first two digits of any temporal motif are always \texttt{01}, to denote that the first event occurred from node \texttt{0} to node \texttt{1}.
~\cref{fig:notation} presents some examples.
For instance, \texttt{0110} denotes a type of $M^2$ that consists of two opposite events occurring between two nodes (\texttt{01} and \texttt{10}).
Similarly, \texttt{011202} represents a type of $M^3$ where the first event is from the black node (\texttt{0}) to the white node (\texttt{1}), the second one is from the white node (\texttt{1}) to the gray node (\texttt{2}), and the last event is from the black node (\texttt{0}) to the gray node (\texttt{2}).
Based on the definition, each digit notation denotes one type of temporal motif, and each type of temporal motif corresponds to a unique digit notation.

\section{Motif Transition Model}\label{sec:mtm}
In this section, we introduce the motif transition model, \textit{MTM}, to generate synthetic temporal networks with realistic global and local characteristics.
Unlike the existing models that generate different types of temporal motifs independently over time~\cite{purohit2018temporal, zeno2021dymond}, the core idea of \textit{MTM} is to model the evolution of temporal motifs as a stochastic process.
For example, a 1-event motif can evolve into a wedge if there comes a new event that has one node in common, or a wedge motif can evolve into a triangle if there is a new event connecting the two nodes that are not directly connected.
We generate temporal synthetic networks by simulating the evolution of temporal motifs.

\subsection{Motif Transition Process}
We use motif transition process to model the evolutions of temporal motifs. Here we givs the definition of motif transition and motif transition process.

\begin{definition}
{\bf (Motif transition)} In a given graph $G$, a motif $M^l_i=(V', E')$ transitions to a motif $M^{l+1}_j=(V'', E'')$
if there is a new event $e_{l+1}=(u_{l+1}, v_{l+1}, t_{l+1})$ such that 

\begin{itemize}\label{def:trans}
\item $V'' = V' \cup \{u_{l+1}, v_{l+1}\} $ and $E'' = E' \cup e_{l+1}$, 
\item $\{u_{l+1}, v_{l+1}\} \cap V' \neq \emptyset$, i.e., the new event is adjacent to $M^l_i$,
\item $t_{l+1} > t_l$, where $t_l$ is the timestamp of the last event in $M^l_i$,
\item $M^l_i$ does not transition to another motif before $e_{l+1}$ arrives, i.e, there does not exist an event $e^*=(u^*, v^*, t^*) \in G$ that satisfies all requirements above and $t^* < t_{l+1}$.
\end{itemize}
We use $\mathcal{T}(M^l_i \rightarrow M^{l+1}_j)$ to denote the transition, and $\{\mathcal{T}(M^l_i \rightarrow M^{l+1}_j)\}$ to denote the set of instances of such transition in the input graph. We define the arrival time of the new event as the transition time: $\Delta_t = t_{l+1} - t_l$.
\end{definition}

\begin{figure}[t!]
\centering
\includegraphics[width=0.95\linewidth]{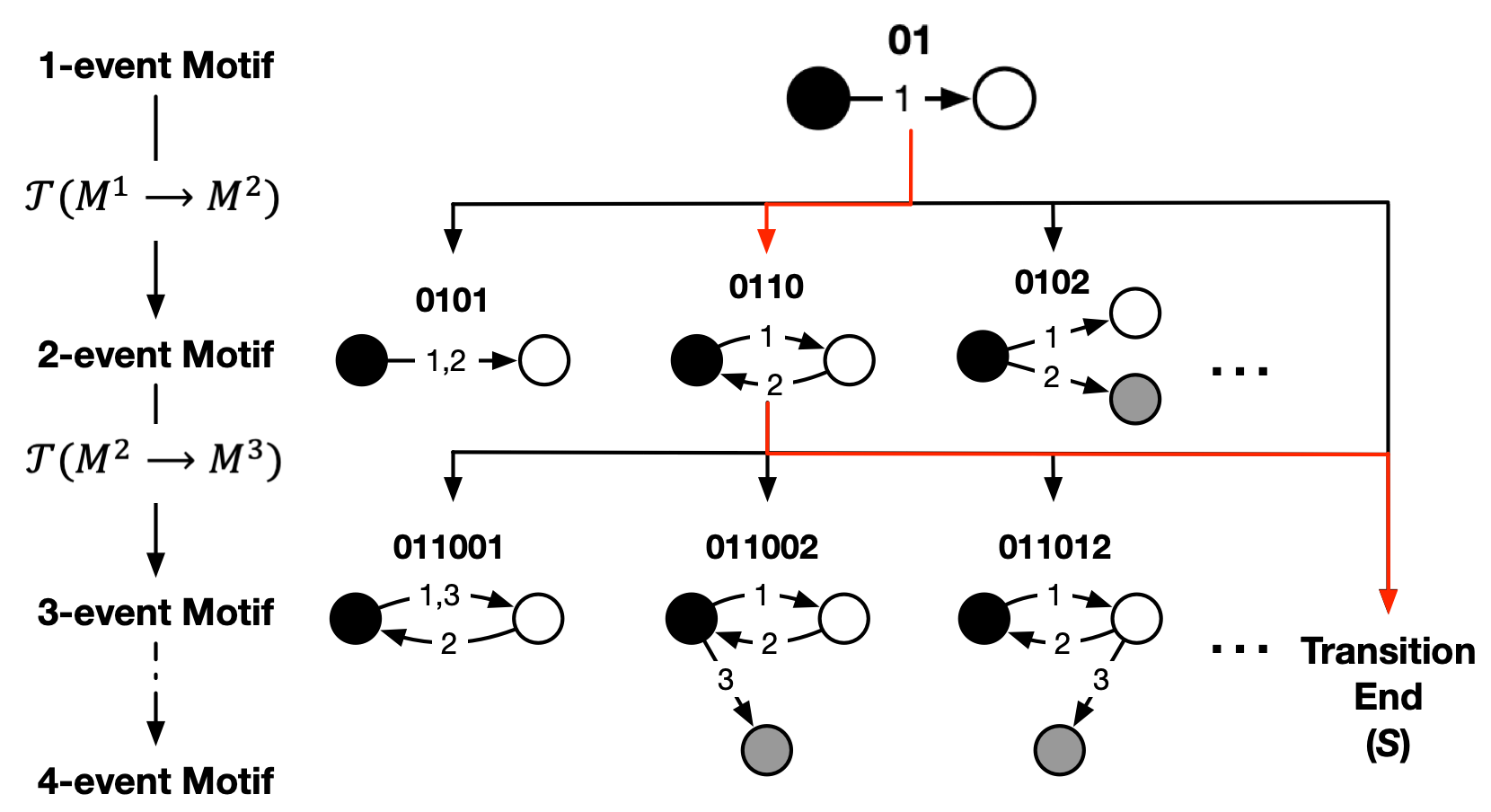}
\vspace{-3ex}
\caption{\small Examples of motif transition processes. $S$ indicates that the transition process ends and no more events can be added to the current motif. A transition process is sequence of motif transitions from an event \texttt{01} to the transition end $S$.
}
\vspace{-3ex}
\label{fig:transition-process}
\end{figure}

\cref{fig:transition-process} gives some examples of the motif transition. 
In our digit notation, the first motif in the transition is always a prefix of the second motif.
For example, a single event (\texttt{01}) can transition into one of the six types of 2-event motifs by the arrival of a consecutive event. Similarly, a 2-event motif, say \texttt{0110}, can transition into a 3-event motif such as \texttt{011001}, \texttt{011002}.

\begin{definition}
{\bf (Motif transition process)} In a given graph $G$, we define the motif transition process as a sequence of motif transitions, $\mathcal{T}(M^1 {\rightarrow} \dots {\rightarrow} M^l_i {\rightarrow} S)$, with respect to the {\bf transition size limit} $l_{\mathrm{max}}$ and {\bf transition time limit} $\delta$. $S$ denotes the stopping state. The motif transition process starts from 1-event motif and ends at $M^l_i$ if either one of the following conditions is true:
\begin{itemize}
\item The size of the $M^l_i$ is equal to the transition size limit, i.e., $l = l_{\mathrm{max}}$,
\item Within the time window $(t_l, t_l + \delta]$, there does not exist a new event $e_{l+1}$ to create the next transition $\mathcal{T}(M^l_i {\rightarrow} M^{l+1}_j)$. 
\end{itemize}
\end{definition}

\begin{figure}[!h]
\vspace{-2ex}
\centering
\includegraphics[width=\linewidth]{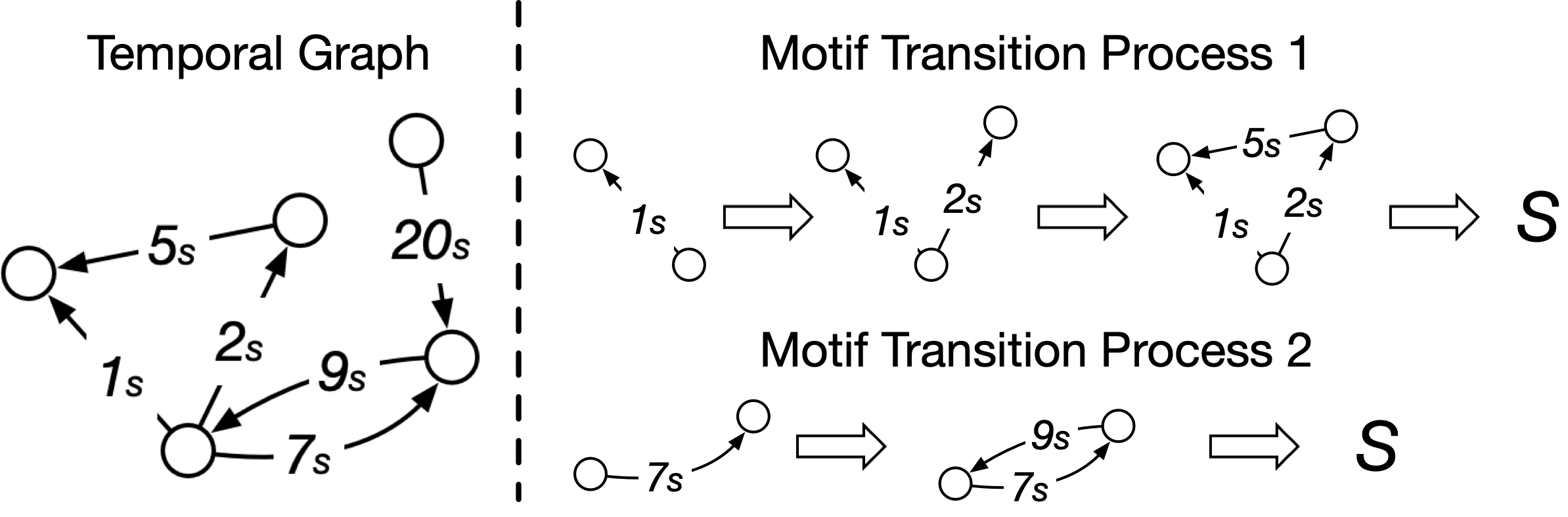}
\vspace{-4ex}
\caption{\small Motif transition processes for $l_{\mathrm{max}}=3$ and $\delta=5$s.}
\vspace{-2ex}
\label{fig:transition-process-example}
\end{figure}

\cref{fig:transition-process-example} gives examples of the motif transition process on a toy graph for $l_{\mathrm{max}}=3$ and $\delta=5$s. 
There is a motif transition process from 1s to 5s which stops as the size of the triangle motif is equal to the transition size limit $l_{\mathrm{max}}$. Another motif transition process starts at 7s and ends at 9s because there does not exist a new event in $(9\mathrm{s}, 14\mathrm{s}]$ to create a new transition.

We categorize events in a temporal network to two classes: {\bf cold events}, which are the first in a motif transition process, and {\bf hot events}, which are subsequent events added in a motif transition process.
We denote the set of cold events as $CE$.
The first event in a temporal graph is always a cold event.
Each cold event can only trigger one motif transition process, and each motif transition process only contains one cold event.
Given the state of the current motif, the motif transition process captures the arrival of the new event, which can be modeled as a Markov process.
Based on this idea, we develop a graph generative model to simulate temporal networks as a stochastic process of new event arrivals.

\subsection{Graph Generative Model}
\begin{figure*}[t!]
\centering
\includegraphics[width=0.9\linewidth]{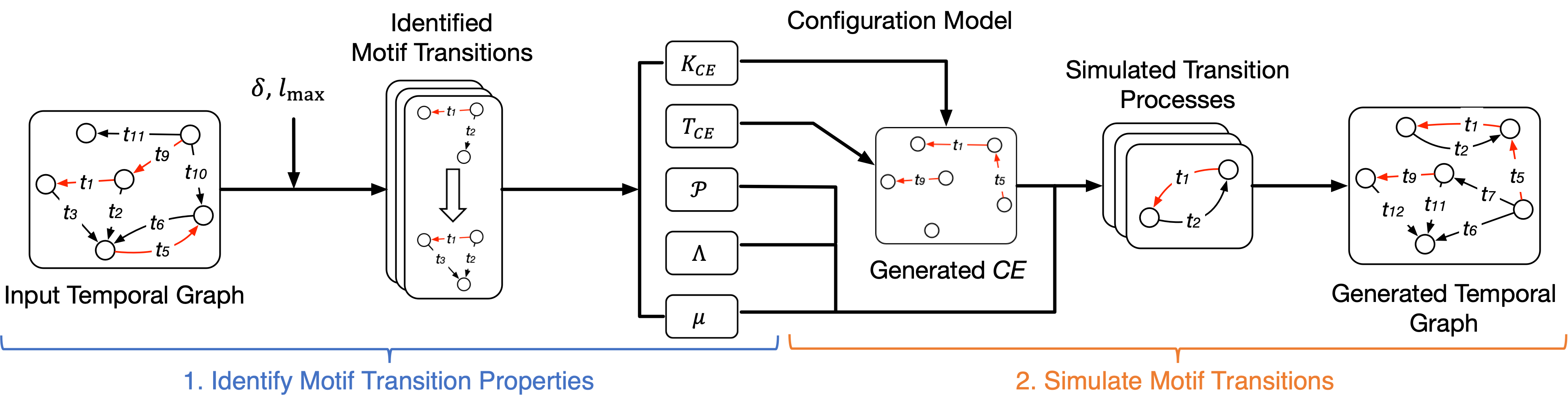}
\vspace{-3ex}
\caption{\small The proposed motif transition model (\textit{MTM}). 
}
\vspace{-2ex}
\label{fig:MTM}
\end{figure*}

We use motif transition processes to develop the motif transition model (\textit{MTM}), a stochastic model to generate synthetic temporal networks.
\cref{fig:MTM} gives an overview of the \textit{MTM} model.
Our model takes a temporal graph as input and generates a synthetic network in two steps: we first identify the motif transition properties of the input graph, and then generate the synthetic network by simulating the motif transition processes. 

\subsubsection{Identifying motif transition properties}
In this step, we calculate the following five properties from the original input graph, which are later used for simulating the motif transitions in the second step. 

\begin{itemize}
\item \textbf{Degree distribution of cold events, $K_{CE}$.}  We create the static projection of the cold events, denoted by $\overline{CE}$, such that $\overline{CE} = \{(u_i, v_i)\ | (u_i, v_i, t_i) \in CE\}$. 
We define $K_{CE}$ as the list of in-degrees and out-degrees of the nodes in $\overline{CE}$. We use $K_{CE}$ to generate the new cold events in the synthetic network.

\item \textbf{Timestamps of cold events, $T_{CE}$.} We identify the timestamps of all the cold events in $G$. We use $T_{CE}$ in generation of the cold events in the synthetic network.

\item \textbf{Motif transition probabilities, $\mathcal{P} = [P_1, P_2, \dots, P_{|\mathbb{T}|}]$}. We identify all motif transitions in the input graph and calculate the conditional probability of each type of motif transitions $\mathcal{T}(M^l_i {\rightarrow} M^{l+1}_j)$:
$$
P(M^{l+1}_j | M^l_i) = \frac{|\{\mathcal{T}(M^l_i {\rightarrow} M^{l+1}_j)\}|}{|\{\mathcal{T}(M^l_i {\rightarrow} S)\}| + \sum_k |\{\mathcal{T}(M^l_i {\rightarrow} M^{l+1}_k)\}|}
$$
where $M^{l+1}_j$ is a type of $(l+1)$-event motif that $M^l_i$ can transition into and $|\{\mathcal{T}(M^l_i {\rightarrow} M^{l+1}_j)\}|$ is the number of corresponding transition instances. $|\{\mathcal{T}(M^l_i {\rightarrow} S)\}|$ is the number of transition processes that end at $M^l_i$, and $\sum_i |\{\mathcal{T}(M^l_i {\rightarrow} M^{l+1}_k)\}|$ is the total number of all motif transition instances from $M^l_i$.
$|\mathbb{T}|$ stands for the total number of transition types under the $l_{\mathrm{max}}$ limit. There are 6 types of $\mathcal{T}(M^1{\rightarrow} M^2)$ transitions, 60 types of $\mathcal{T}(M^2{\rightarrow} M^3)$ transitions, and 888 types of $\mathcal{T}(M^3{\rightarrow} M^4)$ transitions. Threfore, $|\mathbb{T}| = 6$ when $l_{\mathrm{max}}=2$, $|\mathbb{T}| = 66$ when $l_{\mathrm{max}}=3$, and $|\mathbb{T}| = 954$ when $l_{\mathrm{max}}=4$.

\item \textbf{Motif transition rates, $\Lambda = [\lambda_1, \lambda_2, \dots, \lambda_{|\mathbb{T}|}]$.} 
Since the motif transition is nothing but arrival of a new event, inspired by the previous works on stochastic temporal network models~\cite{Masuda16,porter2022analytical},
we use Poisson process to model the transition time $\Delta_t$ (i.e., the arrival time of the new event) for each type of motif transition. In particular, we set the transition rate (i.e., the arrival rate of the new event), denoted as $\lambda(M^l_i {\rightarrow} M^{l+1}_j)$, of a motif transition to the average of the transition times of all the instances of that motif transition: $1 / \mathrm{mean}(\overrightarrow{\Delta_t})$.

\item \textbf{Average number of edges in the motif transition process, $\mu$.} For each motif transition process in the input graph $\mathcal{T}(M^1 {\rightarrow} \dots {\rightarrow} M^l_i {\rightarrow} S)$, we construct the static projection of $M^l_i$ and consider the number of edges in it. Then we calculate the average number of those.

\end{itemize}

\begin{algorithm}[!t]
\small
\caption{\textsc{Motif Transition Model (MTM)}}\label{alg:main}
\hspace*{\algorithmicindent} \textbf{Input:} A temporal graph as an ordered list of events $E$, the transition\\
\hspace{-11ex} time constraint $\delta$, the transition size constraint $l_{\mathrm{max}}$\\
\hspace{-10ex} \hspace*{\algorithmicindent} \textbf{Output:} A synthetic temporal graph as a list of events $E_{\mathrm{out}}$ 
\begin{algorithmic}[1]
\State $E_{\mathrm{out}} \gets \emptyset$
\State $K_{CE}, T_{CE}, \mathcal{P}, \Lambda, \mu \gets $ \textsc{Compute Motif Transitions}$(E, \delta, l_{\mathrm{max}})$ \label{line:MT}
\State $CE' = \textsc{Configuration Model}(K_{CE}, T_{CE})$ \label{line:configuration}

\ForAll{$\{u, v, t\} \in CE'$} \label{line:simulate_0}
\State $E_{\mathrm{out}} \gets E_{\mathrm{out}} \cup \{u, v, t\}$
\State $V' = [u, v]$, $M = $ \texttt{01}
\While {number of events in $M \leq l_{\mathrm{max}}$}\label{line:size}
\State $M' \gets$ randomly choose by using $P(M'|M) \in \mathcal{P}$ \label{line:select_next}
\If {$M' = S$} \label{line:sationary}
\State \textbf{break}
\EndIf
\State digit$_{u'}$, digit$_{v'} \gets$ last two digits of $M'$ \Comment{the new event}
\If{digit$_{u'} > |V'|$} \label{line:new_node_0}
\Comment{request a new edge}
\State $v' = V'[$digit$_{v'}]$
\State randomly choose $u'$ to form $(u',v')$ based on Equation \ref{eq:pa}
\State $V'$.add($u'$)
\ElsIf {digit$_{v'} > |V'|$}
\Comment{request a new edge}
\State $u' = V'[$digit$_{u'}]$
\State randomly choose $v'$ to form $(u',v')$ based on Equation \ref{eq:pa}
\State $V'$.add($v'$) \label{line:pa_1}
\Else
\State $u' = V'[$digit$_{u'}]$
\State $v' = V'[$digit$_{v'}]$
\EndIf

\State $t = t + \mathrm{Pois}(\lambda(M \rightarrow M'))$ \label{line:generate_poisson}
\State $E_{\mathrm{out}} \gets E_{\mathrm{out}} \cup \{u', v', t\}$
\State $M = M'$
\EndWhile

\EndFor
\State \textbf{return} $E_{\mathrm{out}}$ \label{line:simulate_1}
\end{algorithmic}
\end{algorithm}

\subsubsection{Simulate motif transitions}
In the second step, we generate the synthetic network using the motif transition properties captured in the first step.
We utilize $K_{CE}$ and $T_{CE}$ to generate the cold events, and $\mathcal{P}$, $\Lambda$, $\mu$ to generate the hot events.
We first employ the configuration model~\cite{newman2001random} to generate static edges from the degree distribution $K_{CE}$, and then apply weight-constrained link shuffling~\cite{gauvin2018randomized} to randomly assign the timestamps $T_{CE}$ on the edges to create the cold events  for the output graph, denoted by $CE'$.

After generating the cold events, $CE'$, we process each in the chronological order to generate the hot events.
For each cold event, we randomly generate a motif transition process using the transition probabilities of the input graph, $\mathcal{P}$. 
We store all the cold and hot events in $E_{\mathrm{out}}$.
We simulate the timestamps of the new events by the arrival times generated from the Poisson process $\mathrm{Pois}(\lambda)$. 
In particular, we generate the timestamp of a new event as $t_{l+1} = t_l + \Delta_t$, where $P(\Delta_t > x) = e^{-\lambda x}$.
We continue adding new events to the transition process until it reaches the stopping state $S$.
After all cold events have been processed, the model gives the event list $E_{\mathrm{out}}$ as the output graph.

A new event may create an edge that does not exist in the previous transition.
For example, the transition from \texttt{01} to \texttt{0102} brings a new event which does not occur on an edge in the current motif.
We consider two ways to select the edge $(u', v')$ for the new event: (1) we randomly select an edge that is in the static projection of the current output event list, i.e., $(u', v') \in \overline{E_{\mathrm{out}}}$; or (2) we randomly create a new edge that is not in $\overline{E_{\mathrm{out}}}$, which will increase the size of $\overline{E_{\mathrm{out}}}$ by one. We set the probability to create a new edge as
\begin{equation} \label{eq:pa}
p((u', v') \notin \overline{E_{\mathrm{out}}}) = \frac{|\overline{E}| - |\overline{CE'}|}{(\mu -1) |CE'|}
\end{equation}
where $\overline{CE'}$ is its static projection of cold events for the output graph, $\overline{E}$ is the static projection of the input graph, and $\mu$ is the average number of edges in motif transition processes. Since each cold event will transition into a motif with $\mu$ edges on average, the transition processes will request $(\mu -1) |CE'|$ new edges in addition to the cold events. For each request we create a new edge based on the probability in Equation~\ref{eq:pa}, which will fill the difference between the number of edges in the input graph and the cold events $|\overline{E}| - |\overline{CE'}|$.

\begin{algorithm}[!t]
\small
\caption{\textsc{Compute Motif Transitions}}\label{alg:transition}
\hspace*{\algorithmicindent} \textbf{Input:} A temporal graph as an ordered list of events $E$, the transition\\
\hspace{-11ex} time constraint $\delta$, the transition size constraint $l_{\mathrm{max}}$\\
\hspace{-2ex} \textbf{Output:} Degrees and timestamps of cold events $K_{CE}$ and $T_{CE}$,\\
\hspace{0ex} transition probabilities $\mathcal{P}$, transition rates $\Lambda$, average number of\\
\hspace{-22ex} edges in the motif transition processes $\mu$
\begin{algorithmic}[1]
\State $CE \gets \emptyset$ \Comment{set of cold events}
\State $L \gets [ ]$ \Comment{list of number of edges in motif transition processes}
\State ${\mathcal{T}_{\mathrm{active}}} \gets [ ]$ \Comment{list of active transition processes}
\State $\overrightarrow{\Delta_t} \gets [[ ]]$ \Comment{list of list of transition times}
\State $C \gets [0]$ \Comment{list of \# of instances for each motif transition (initially 0)}

\ForAll{$(u, v, t) \in E$} \label{line:identify_transition_0}
\State $ce$ = \textsc{True} \Comment{cold event flag}
\ForAll{$\mathcal{T} \in {\mathcal{T}_{\mathrm{active}}} $}
\State $M \gets$ the last motif of $\mathcal{T}$
\State $t_{\mathrm{max}} \gets$ the last timestamp of $M$
\State $V' \gets$ all the nodes in $M$
\If {size of $M \geq l_{\mathrm{max}}$ or $t - t_{\mathrm{max}} > \delta$}
\State ${\mathcal{T}_{\mathrm{active}}}$.remove $(\mathcal{T})$
\Comment{$\mathcal{T}$ ends}
\State $L$.add (number of edges in $M$)
\ElsIf {$\{u,v\} \cap V' \neq \emptyset$} \label{line:check_intersect}
\State $M' \gets M \cup (u, v, t)$
\Comment{add the new event }
\State $C[M \rightarrow M'] = C[M \rightarrow M'] + 1$ \label{line:count}
\State $\overrightarrow{\Delta_t}[M \rightarrow M']$.add ($t - t_{\mathrm{max}})$ \label{line:time}
\State ${\mathcal{T}_{\mathrm{active}}}$.remove $(\mathcal{T})$
\State ${\mathcal{T}_{\mathrm{active}}}$.add $(\mathcal{T} + M')$ \label{line:prefix_update_0}
\Comment{update the list of active transitions}
\State $ce = $ \textsc{False} \label{line:not_initial}
\EndIf
\EndFor

\If {$ce =$ \textsc{True}}
\State $CE \gets CE \cup (u, v, t)$ \Comment{add to cold events}
\State $M' \gets [(u, v, t)]$
\EndIf
\State ${\mathcal{T}_{\mathrm{active}}}$.add$ (M')$ \label{line:prefix_update_1}
\Comment{update the set of active transitions}
\EndFor
\ForAll{$\mathcal{T} \in {\mathcal{T}_{\mathrm{active}}}$}
\State $L$.add (number of edges in the last motif of $\mathcal{T}$)
\EndFor

\ForAll{$C[M \rightarrow M'] \in C$}\label{line:calculate_0}
\State $P(M'|M) \gets$ calculate transition probability using $C[M \rightarrow M']$
\State $\lambda(M \rightarrow M') \gets$ mean$(1 / \Delta_t[M \rightarrow M'])$
\State $P(M \rightarrow S) =  1- \sum(P(M'|M))$\label{line:calculate_1}
\EndFor

\State $\mathcal{P} \gets$ list of $P$
\State $\Lambda \gets$ list of $\lambda$
\State $K_{CE}, T_{CE} \gets \textsc{Get degrees and timestamps}~(CE)$
\State $u \gets$mean $(L)$
\State \textbf{return} $K_{CE}$, $T_{CE}$, $\mathcal{P}$, $\Lambda$, $\mu$
\end{algorithmic}
\end{algorithm}

\subsubsection{MTM algorithm}

Algorithm \ref{alg:main} provides the pseudocode of the motif transition model. We first calculate the motif transitions in Line \ref{line:MT} using Algorithm \ref{alg:transition}. Then we simulate the cold events in \cref{line:configuration}, and generate motif transitions from each cold event until the size reaches the transition size limit (Line \ref{line:size}) or reaches to the stopping state $S$ (Line \ref{line:sationary}). At each step, we randomly select the next transition according to the calculated transition probabilities (Line \ref{line:select_next}), and generate the arrival time of the new event from Poisson distribution (Line \ref{line:generate_poisson}). If the transition process requests a new edge, we randomly select it using \cref{eq:pa} (Line \ref{line:new_node_0} to \ref{line:pa_1}). The simulation process (Line \ref{line:simulate_0} to \ref{line:simulate_1}) is a linear algorithm with $O(|CE| \cdot l_{\mathrm{max}})$ runtime complexity as the transitions of cold events are limited by $l_{\mathrm{max}}$. 

Algorithm \ref{alg:transition} shows the pseudocode for calculating the motif transitions.
We maintain a list of active transitions, which are the motif transition processes that can be extended by a new event.
For each event, we check all the active transitions to see if the event can be added to an existing transition (Line \ref{line:check_intersect}).
If so, we record the transition count and the transition time (Lines \ref{line:count} and \ref{line:time}).
Note that if the event is added to an existing motif, it cannot be a cold event (Line \ref{line:not_initial}). Before proceeding to the next event in the list, we update the set of active transitions (Line \ref{line:prefix_update_0} and \ref{line:prefix_update_1}). After getting all the transition counts and transition times, we calculate the transition probabilities $\mathcal{P}$ and the transition rates $\Lambda$ (Line \ref{line:calculate_0} to \ref{line:calculate_1}). 
Note that Algorithm \ref{alg:transition} does not explicitly count the motifs.
Identifying the motif transitions (Line \ref{line:identify_transition_0} to \ref{line:prefix_update_1}) takes $O(|E| \cdot \left|{\mathcal{T}_{\mathrm{active}}}\right|)$ time.
Since each cold event can only trigger one active transition, the number of active transitions, $\left|{\mathcal{T}_{\mathrm{active}}}\right|$, at any step is no greater than the number of cold events, $|CE|$, in any $l_{\mathrm{max}} \cdot \delta$ time window.
The average number of $|CE|$ in any time window is $|CE| \frac{l_{\mathrm{max}} \cdot \delta}{|T|}$, where $|T|$ is the timespan of the input data.
Once the transitions are identified, calculating the $\mathcal{P}$ and $\Lambda$ (Line \ref{line:calculate_0} to \ref{line:calculate_1}) takes constant time as there are a fixed number of transition types.
Therefore, the time complexity of Algorithm \ref{alg:transition} is $O(|E| \cdot |CE| \frac{l_{\mathrm{max}} \cdot \delta}{|T|})$. 
In total, time complexity of {\it MTM} becomes $O(|E| \cdot |CE| \frac{l_{\mathrm{max}} \cdot \delta}{|T|}) + |E|)$. Note that in practice, the time window $l_{\mathrm{max}} \cdot \delta$ is 1,000 to 16,000 times smaller than the entire timespan of the network, $|T|$, and the fraction of cold events is less than 4.5\% of all events.
{\it MTM} requires constant space to store the transition properties, and $\left|{\mathcal{T}_{\mathrm{active}}}\right|$ is a constant at each step. Therefore, the space complexity of {\it MTM} is $O(|CE| + |E|) = O(|E|)$.

\section{Experiments}
In this section, we perform experiments to evaluate our model and compare its performance against several baseline models on various real-world networks. We first investigate the motif transition properties for different $l_{\mathrm{max}}$ and $\delta$ values. Then we evaluate the performance of \textit{MTM}  in three aspects: (1) the ability of preserving the global statistics of the original input graph, (2) the ability of preserving the temporal motifs structures in the original input graph, and (3) and the scalability with respect to the size of the input graph and the model parameters.

\subsection{Setup}
All experiments are performed on a Linux operating system running on a machine with Intel(R) Xeon(R) Gold 6130 CPU processor at 2.10 GHz with 128 GB memory.
We also use NVIDIA V100 16GB GPU to run one of the baseline methods (\textit{TagGen} model~\cite{zhou2020data}).
We implemented {\it MTM} in C++.
{\bf The code is available at \url{https://github.com/erdemUB/KDD23-MTM}.}

\subsubsection{Datasets}
We evaluate {\it MTM} on several real-world temporal networks from various domains, including \texttt{CollegeMsg}, \texttt{Email-EU}, \texttt{Email-EU*}, \texttt{FBWall}, \texttt{SuperUser}, and \texttt{StackOverflow}. The details and statistics are given in Appendix B.

\subsubsection{Baselines}

\begin{figure}[!t]
\begin{subfigure}{0.51\linewidth}
\includegraphics[width=\linewidth]{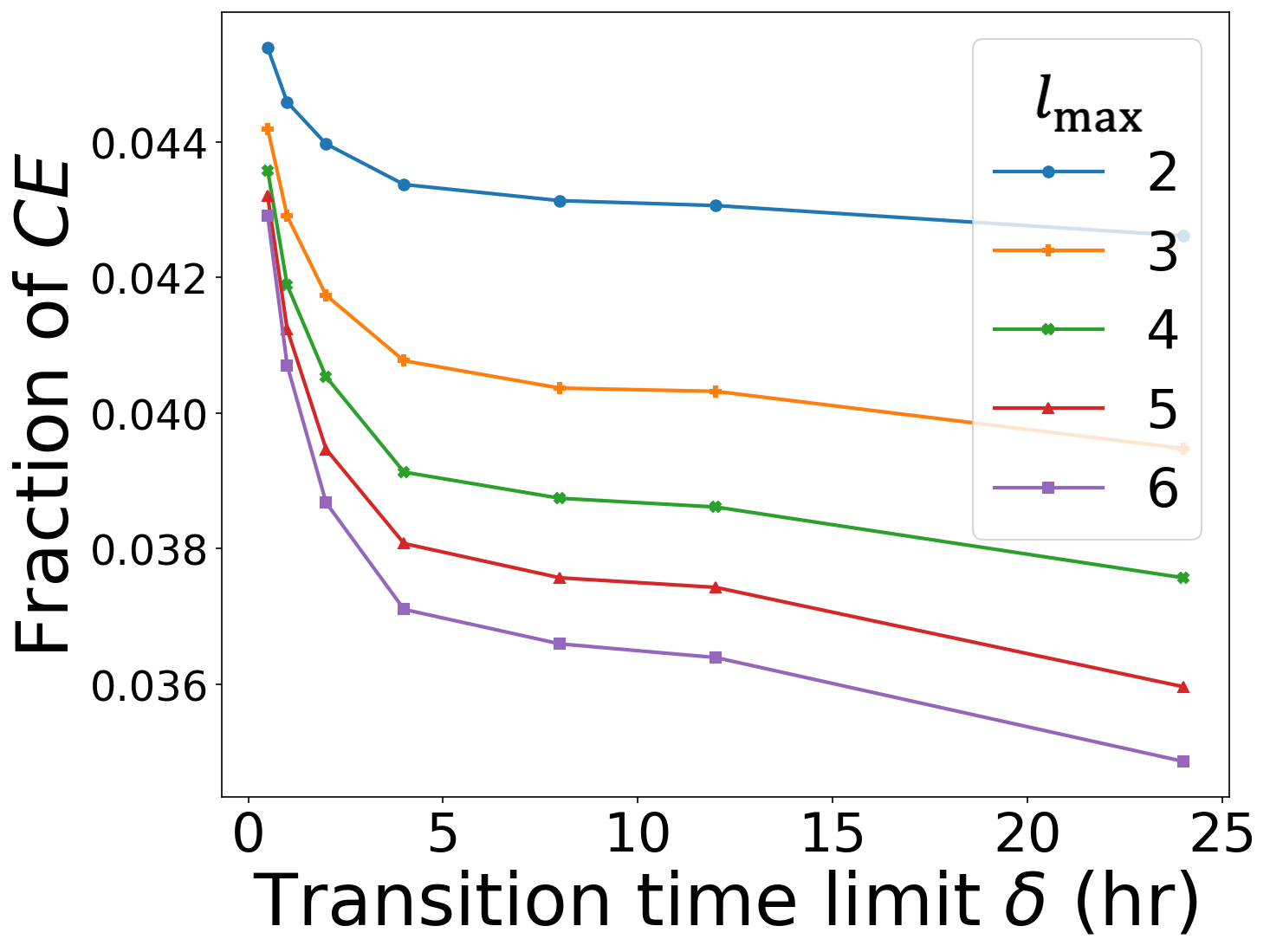}
\caption{Fraction of cold events}
\label{fig:parameter-CE}
\end{subfigure}
\begin{subfigure}{0.48\linewidth}
\includegraphics[width=\linewidth]{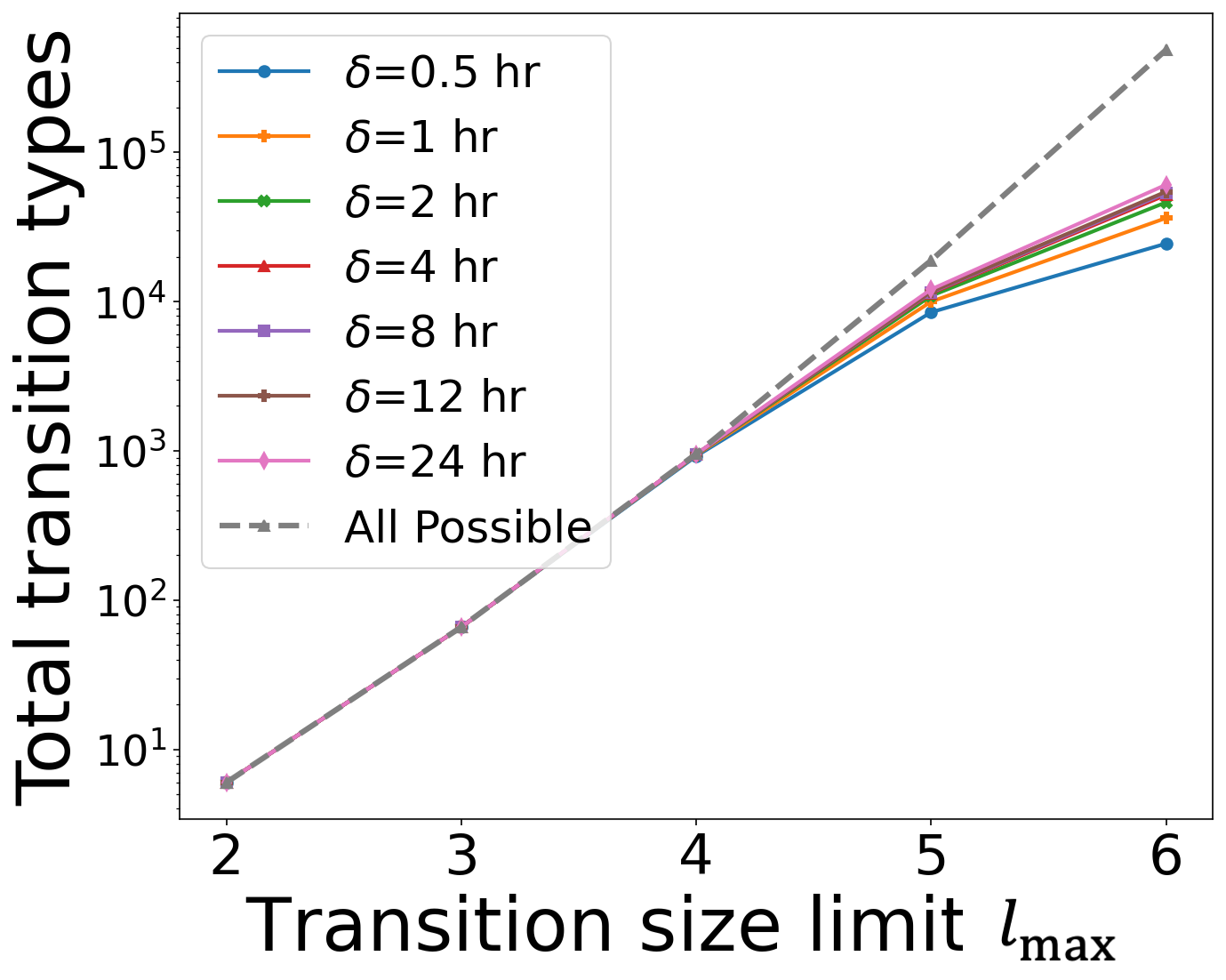}
\caption{Number of transition types}
\label{fig:parameter-P}
\end{subfigure}

\vspace{-2ex}
\caption{\small The impact of the transition size limit ($l_{\mathrm{max}}$) and the transition time limit ($\delta$). \cref{fig:parameter-CE} gives the fraction of the cold events in the input graph. \cref{fig:parameter-P} shows the total number of transition types identified in the real-world network (straight lines) and all possible types of transitions (dashed line).}
\label{fig:parameter}
\vspace{-3ex}
\end{figure}

\begin{figure*}[!t]
\begin{subfigure}{0.24\linewidth}
\includegraphics[width=\linewidth]{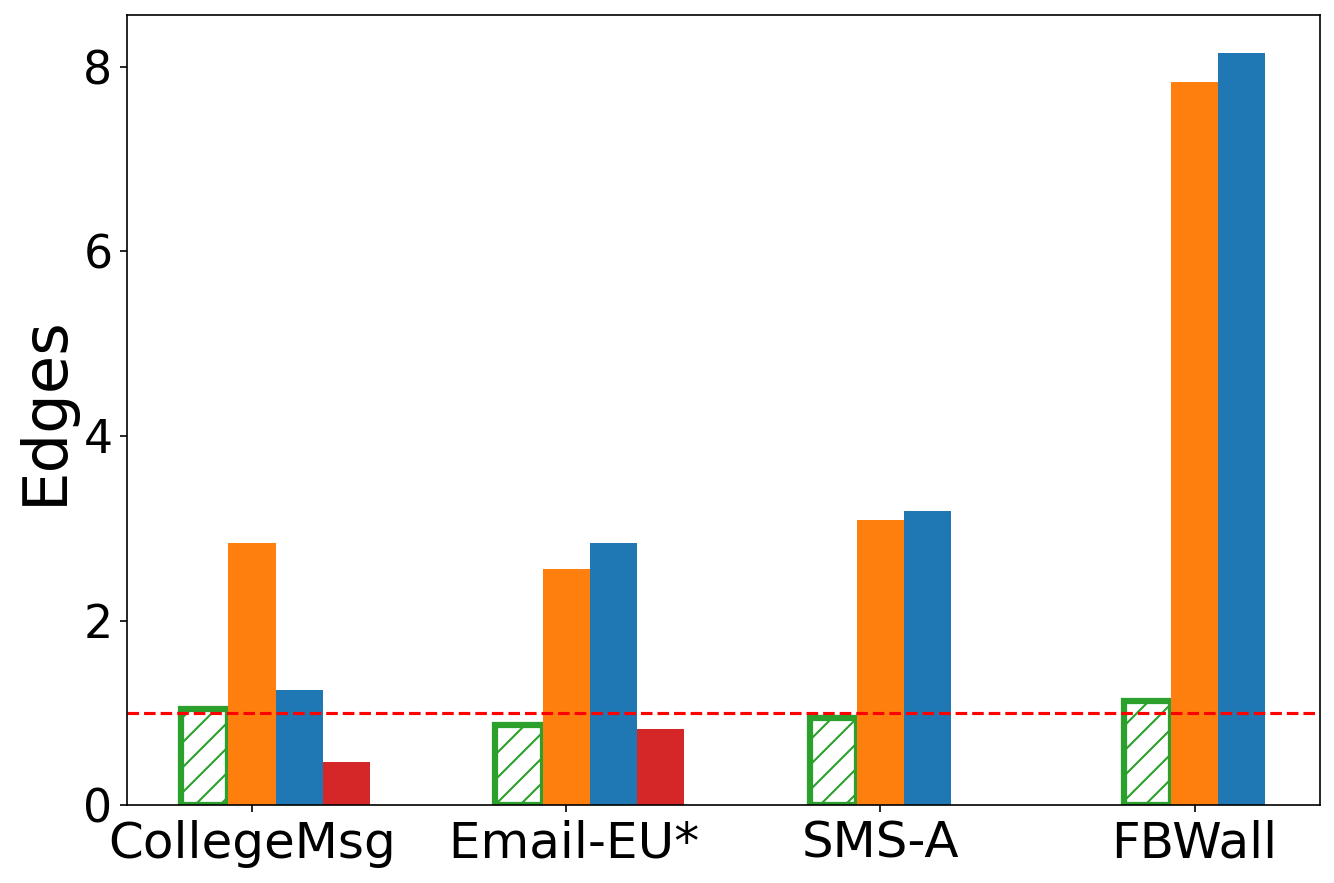}
\caption{Number of edges}
\end{subfigure}
\begin{subfigure}{0.24\linewidth}
\includegraphics[width=\linewidth]{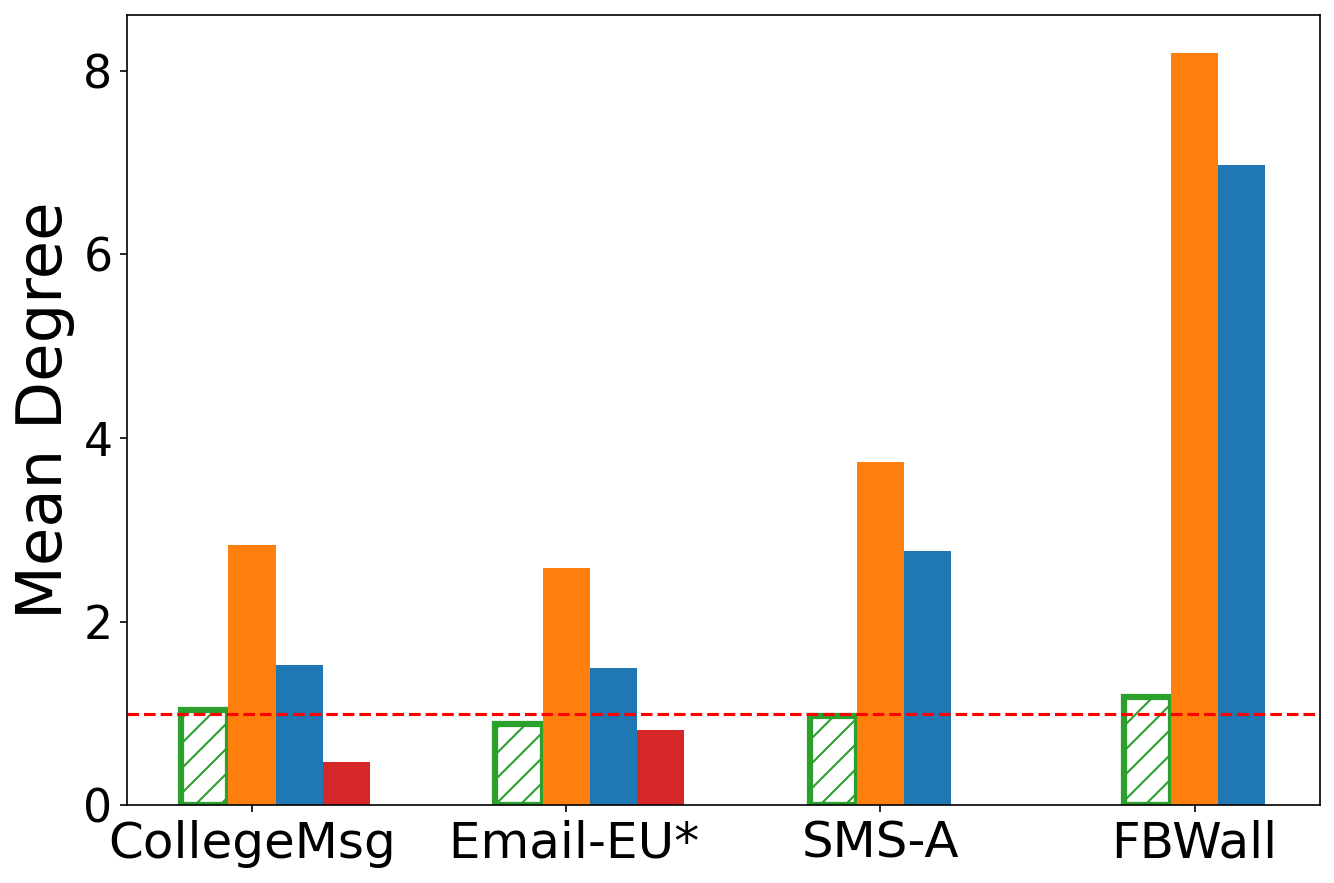}
\caption{Mean Degree}
\end{subfigure}
\begin{subfigure}{0.24\linewidth}
\includegraphics[width=\linewidth]{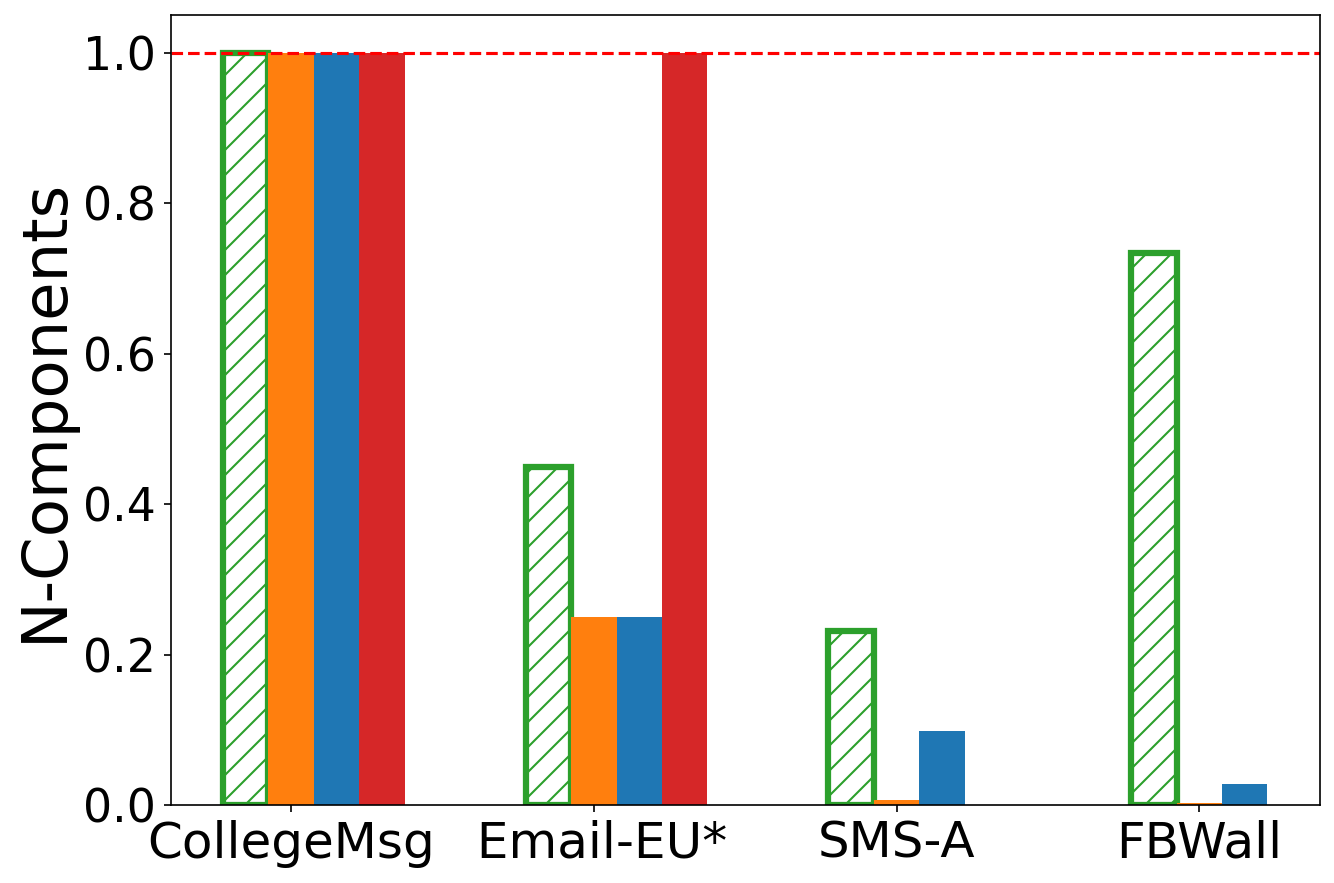}
\caption{N-Components}
\end{subfigure}
\begin{subfigure}{0.24\linewidth}
\includegraphics[width=\linewidth]{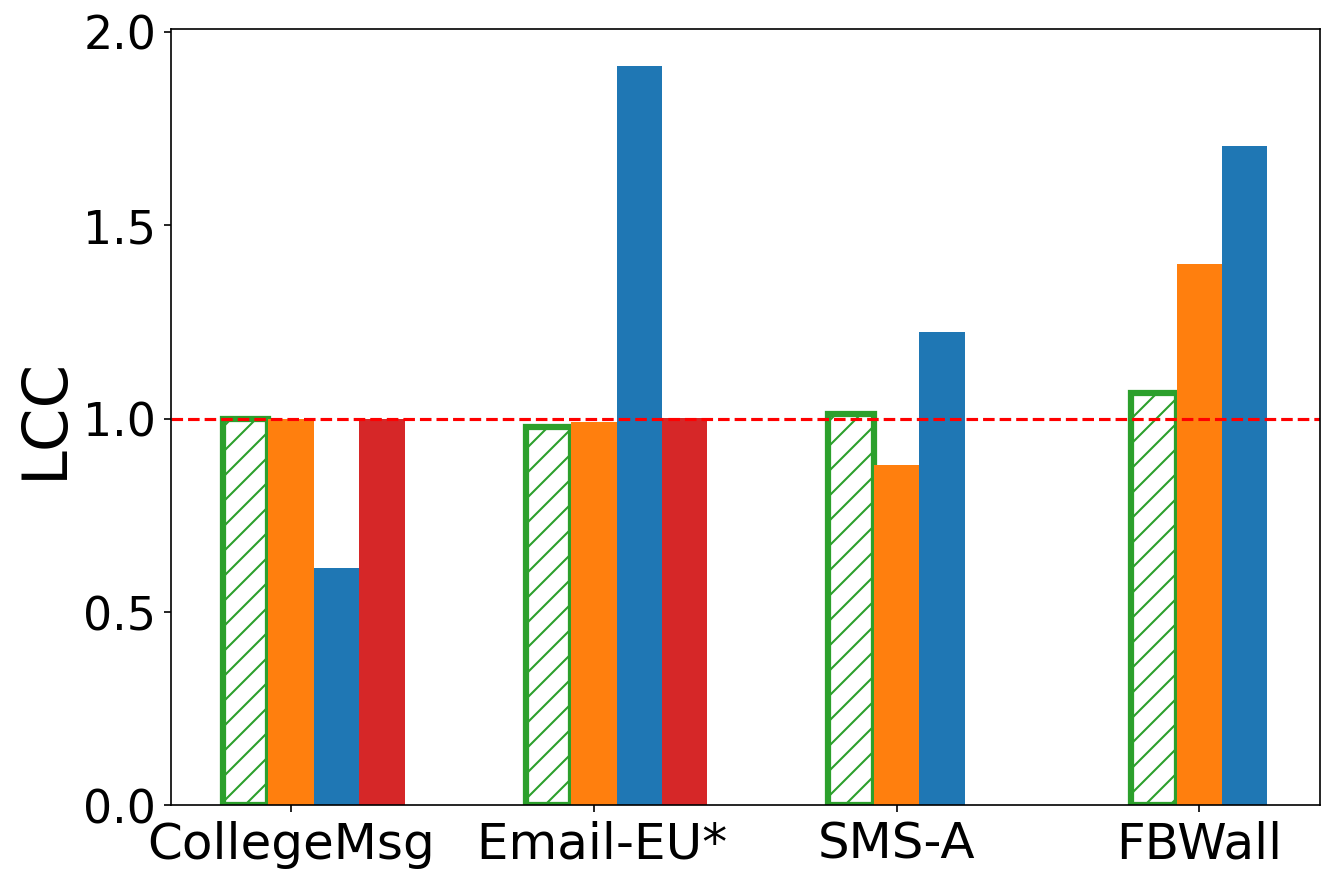}
\caption{LCC}
\end{subfigure}

\begin{subfigure}{0.24\linewidth}
\includegraphics[width=\linewidth]{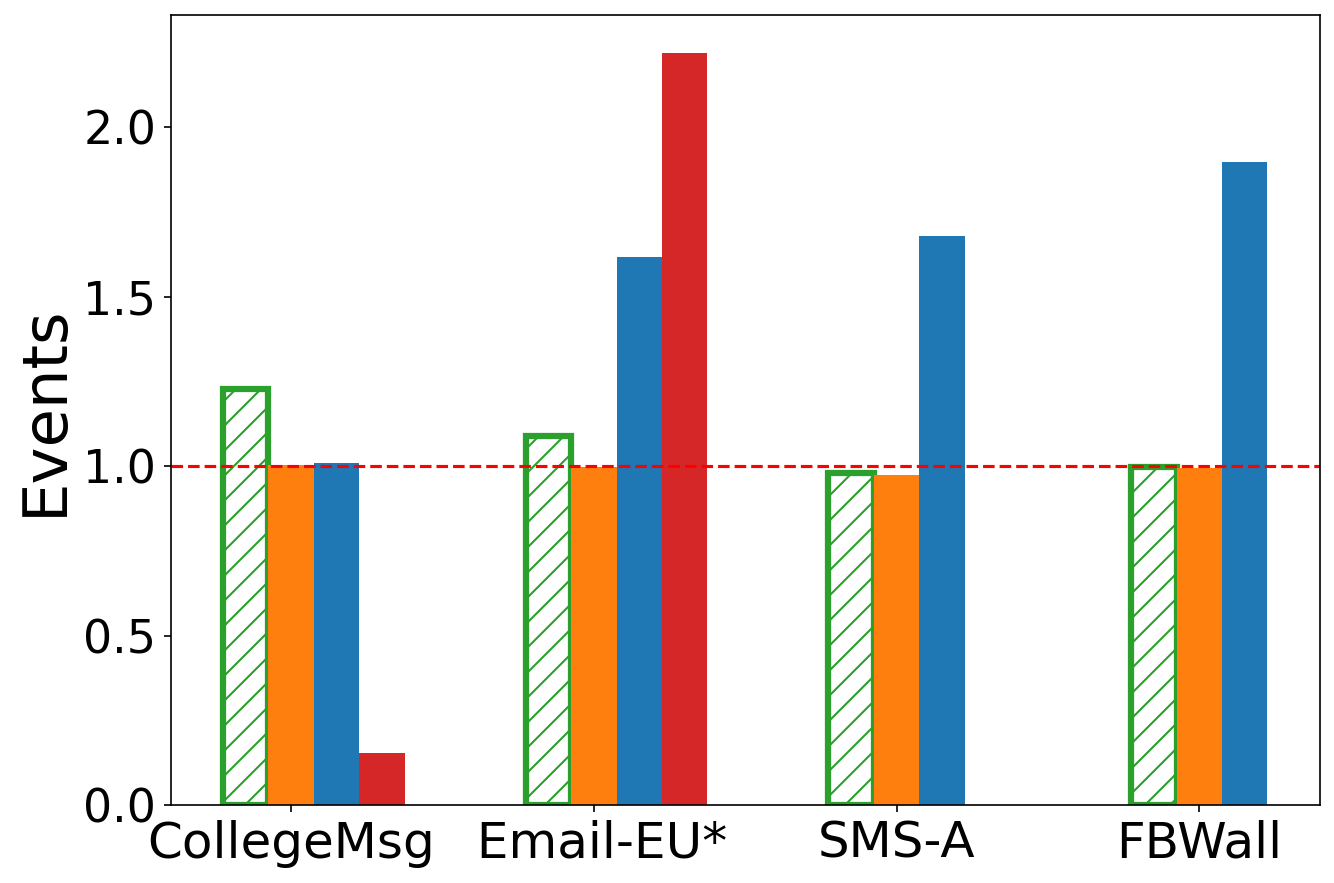}
\caption{Number of events}
\end{subfigure}
\begin{subfigure}{0.24\linewidth}
\includegraphics[width=\linewidth]{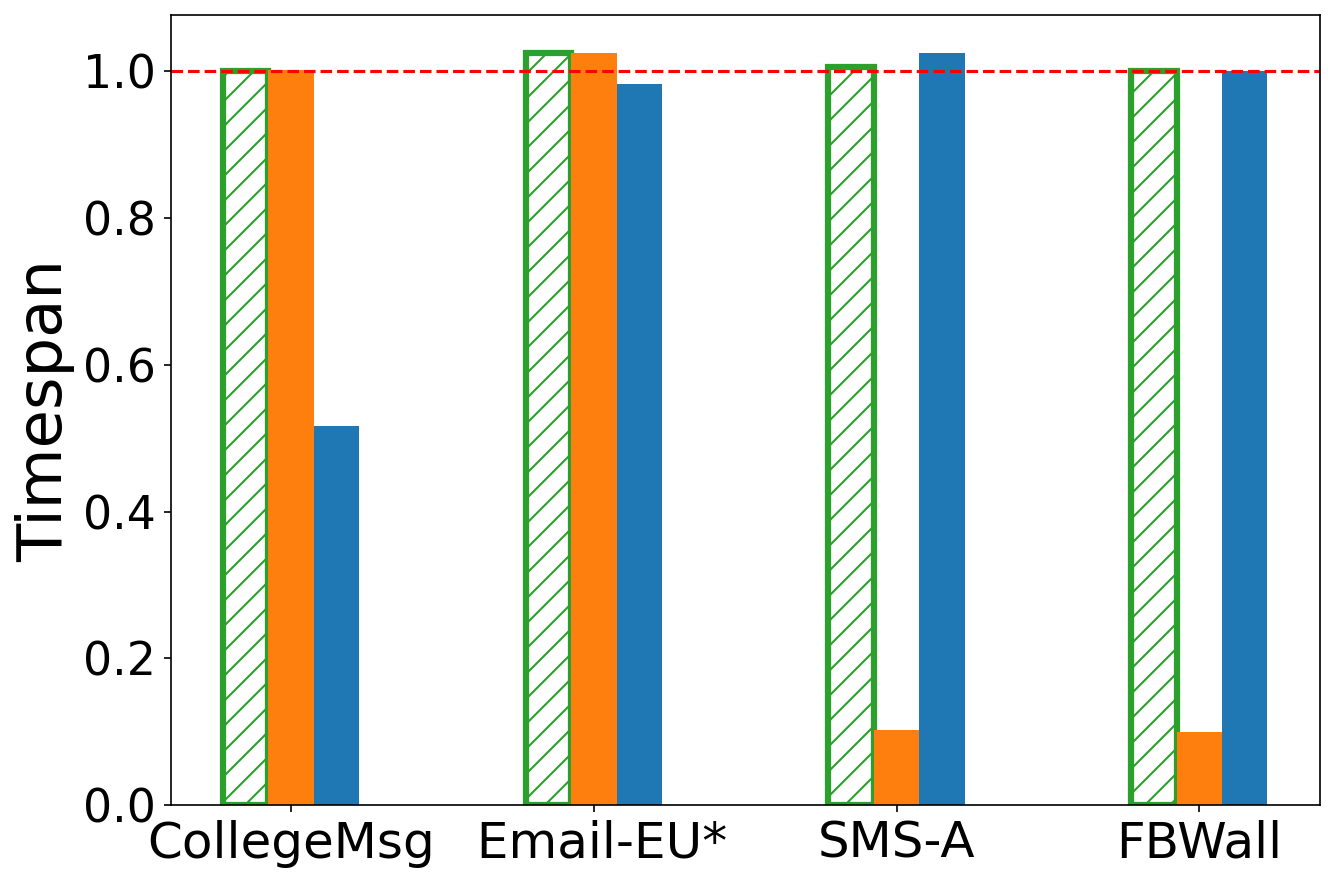}
\caption{Timespan}
\end{subfigure}
\begin{subfigure}{0.24\linewidth}
\includegraphics[width=\linewidth]{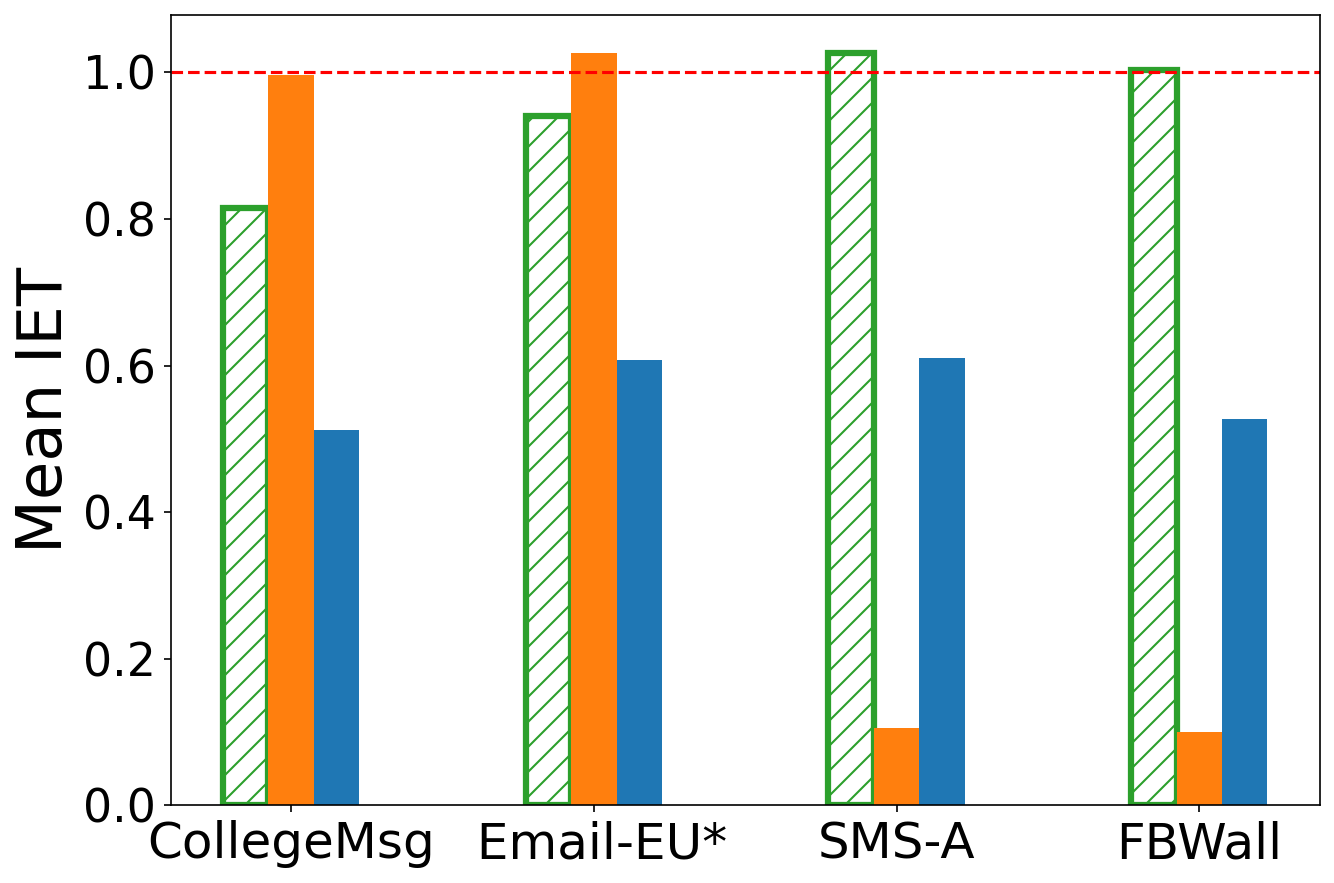}
\caption{Mean IET}
\end{subfigure}
\begin{subfigure}{0.24\linewidth}
\includegraphics[width=\linewidth]{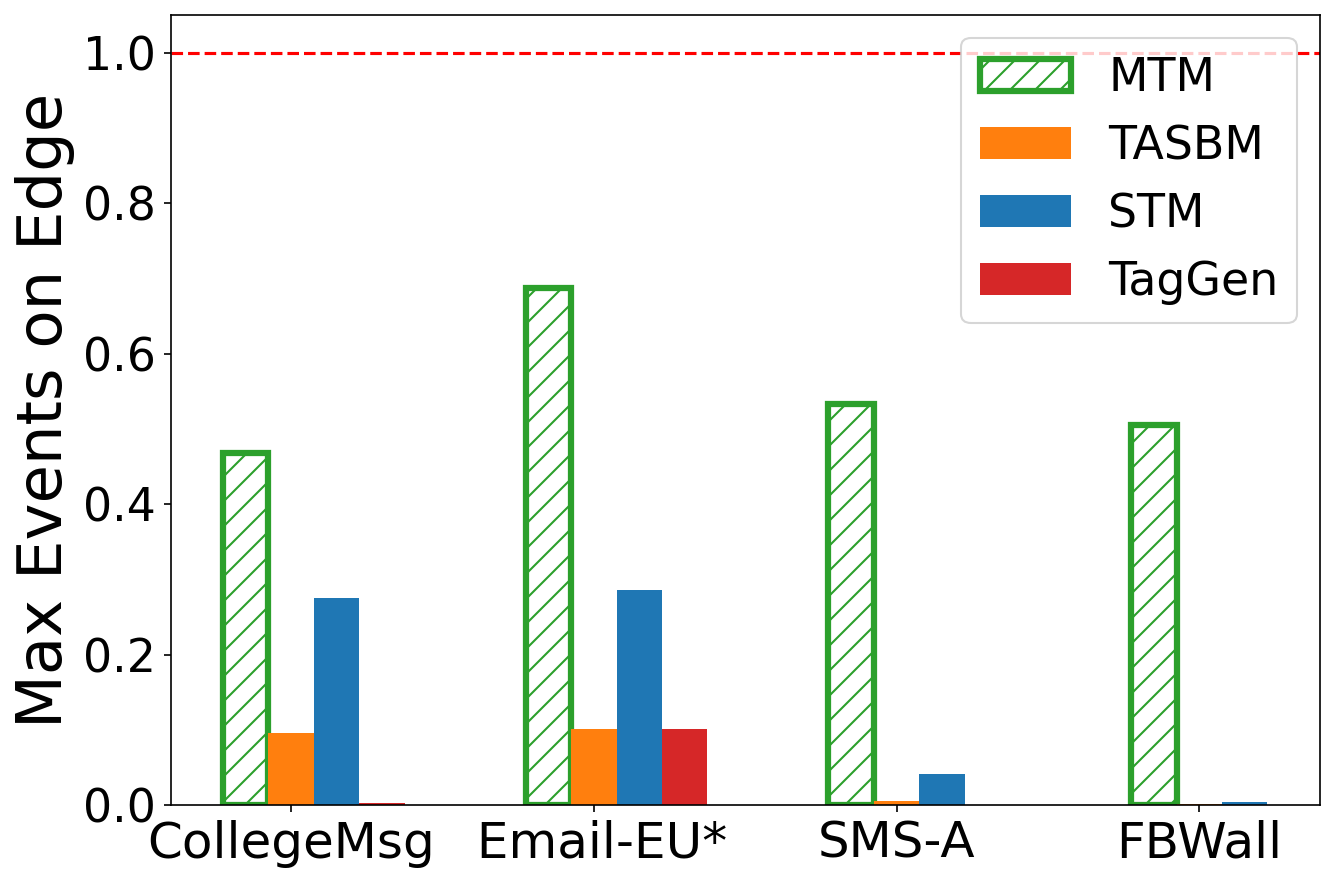}
\caption{Maximum events on edge}
\end{subfigure}
\vspace{-2ex}
\caption{\small Comparison of how the models preserve the global graph statistics. For each graph statistic, we show the ratio of the synthetic networks to the original input graphs. The horizontal dashed line indicates the value of the input graph (ratio=1). The \textit{TagGen} does not generate results for \texttt{SMS-A} and \texttt{FBWall} because both datasets have more than 40K nodes which requires more than 384GB memory for allocation. It also gives networks with very low timespan and mean IET as it generates output graph as a few snapshots, which is why the red bars are not visible in (f), (g), and (h).
}
\label{fig:graph-stats}
\vspace{-3ex}
\end{figure*}

Different from the previous works that model the static motifs in the snapshots of a temporal network~\cite{zeno2021dymond}, our study focuses on truly temporal networks where each event has a unique timestamp.
For this purpose, we select two two baseline models, \textit{TASBM}~\cite{porter2022analytical} and \textit{STM}~\cite{purohit2018temporal}, which process the temporal networks in their original forms. 
In addition, we consider the \textit{TagGen}~\cite{zhou2020data}, a deep generative framework that uses  temporal random walks.

We use the implementation of \textit{TASBM}\footnote{https://github.com/aporter468/motifsanalyticalmodel} (in C++) and \textit{STM}\footnote{https://github.com/temporal-graphs/STM} (in Apache Spark 2.3.0, GraphFrame 0.7.0, and Scala 2.11.8) provided by the authors. We set the number of time windows to 10 for \textit{TASBM} as suggested by the paper. 
We set $\alpha = 1$ for \textit{STM}, as we do not assume to have domain knowledge of the input graphs.
Note that \textit{TagGen} considers temporal networks as a sequence of snapshots by degrading the resolution of the original data. 
Based on the experiment setups described in the \textit{TagGen} paper, we convert the \texttt{CollegeMsg} to 28 snapshots and the \texttt{Email-EU*} to 26 snapshots, while the original data has 58157 and 31750 unique timestamps respectively. We use the available implementation of \textit{TagGen}\footnote{https://github.com/davidchouzdw/TagGen} which creates the graph as a $|V| \times |V| \times |T|$ tensor. As the implementation requires 384GB memory for an input graph with 40K nodes and 30 snapshots, we cannot execute the \textit{TagGen} model on the other datasets with larger size.

\subsection{Impact of Model Parameters}

We first investigate the impact of the model parameters, $l_{\mathrm{max}}$ and $\delta$. In particular, we compare the number of cold events and the number of transition types using different parameters. \cref{fig:parameter} shows the results for \texttt{Email-EU}, and we observe similar patterns in other datasets. 
Using a larger $l_{\mathrm{max}}$ and $\delta$ allows more hot events to be added to the transitions, which leads to less cold events, as shown in~\cref{fig:parameter-CE}.
The decrease in the number of cold events slows down as the $l_{\mathrm{max}}$ and $\delta$ increase.
The overall number of the cold events is small compared to the total number of events in the data, ranging from 3.5\% to 4.5\%. 
As the transition size limit ($l_{\mathrm{max}}$) increases, the possible types of transitions increases exponentially, as shown in~\cref{fig:parameter-P}. However, we do not identify many higher-order transitions in the real-world networks when using a large $l_{\mathrm{max}}$ value. For example, there are more than 466K types of possible $\mathcal{T}(M^5_i {\rightarrow} M^{6}_j)$ transitions, but we only identify 48K in the \texttt{Email-EU} data. Changing the transition time limit, $\delta$, does not have a strong impact on the total number of transition types.
Based on our observations, we set $l_{\mathrm{max}} = 4$ and $\delta = 1$ hour as we achieve less benefits for larger parameters.

\vspace{-1ex}
\subsection{Preserving Global Graph Properties}
Here we examine how well the \textit{MTM} preserves the general graph statistics. 
For each dataset we generate 10 synthetic networks using the \textit{TASBM}, \textit{STM}, and \textit{MTM}, and take the average values for each model. 
While each generated output is a completely different temporal graph, overall we do not observe significant variance between synthetic graphs generated by the same model.
Inspired by the previous models~\cite{purohit2018temporal,zhou2020data,porter2022analytical}, we evaluate four structural global graph metrics: the number of edges, the mean degree (sum of in- and out-degree), the number of connected components (N-Components), and the size of the largest connected components (LCC).
Besides, we consider four temporal graph statistics: the number of events, the entire timespan of the network, the mean inter-event time, and the maximum number of events on an edge.
These metrics are also examined in the previous studies on temporal networks~\cite{vazquez2006modeling,malmgren2008poissonian,karsai2012universal,Masuda16}.
\cref{fig:graph-stats} presents the ratio of synthetic graph statistics (bars) to the input graphs (dashed red line).
As mentioned in the previous section, the \textit{TagGen} model does not apply to \texttt{SMS-A} and \texttt{FBWall} which has more than 40k nodes.

Overall, \textit{MTM} performs the best on preserving the global graph statistics of the input graphs. 
Except the number of components and the maximum events per edge metrics, \textit{MTM} gives synthetic networks with less than 5\% difference than the original graph. 
\textit{TASBM} and \textit{STM} yield synthetic networks with 2 to 8 times more edges than the original graphs, whereas the \textit{TagGen}  gives 50\% less edges and degrees for the \texttt{CollegeMsg} data.
While our model gives more accurate number of components than \textit{TASBM} and \textit{STM}, the \textit{TagGen} performs the best. 
We also observe that \textit{STM} tends to overamplify the size of the largest connected component.

Regarding the temporal graph statistics, \textit{MTM} yields synthetic networks with more accurate characteristics than the baseline models.
The error of the mean IET (inter-event time) is less than 20\% for {\it MTM}.
Although all the models perform poor in preserving the maximum number of events on edges, \textit{MTM} gives synthetic networks with maximum events on edge significantly closer to the original graph.
The reason is that \textit{TASBM} generates events between two nodes purely based on the activity level and the arrival rates, without considering the correlations between the events on the same edge. \textit{STM} only captures the repetitions of events through \texttt{0101} motifs. \textit{TagGen}  processes the original data as snapshot sequences, thus cannot capture the temporal characteristics. Our model, {\it MTM}, on the other hand, considers the complete motif spectrum under the given event limits, which enables to capture the repetitions in any step of the transition process. 
\begin{table}[!b]
\vspace{-2ex}
\captionsetup{justification=centering}
\caption{\small Kolmogorov-Smirnov (KS) statistics to compare the distributions in original and synthetic networks (lower is better).
}
\vspace{-1ex}
\small
\setlength\tabcolsep{1.5pt}
\centering
\begin{tabular}{|l|l|r|r|r|r|}
\hline
Data       & Model & In-degree & Out-degree & IET   & Timestamp \\ \hline
\multirow{3}{*}{CollegeMsg} & \textit{TASBM} & 0.311     & 0.269      & 0.435 & 1.000     \\
 & \textit{STM}   & 0.261     & 0.398      & 0.525 & 1.000     \\
 & \textit{TagGen}   & 0.224     & 0.283      & 0.984 & 1.000     \\
 & \textit{MTM}   & \textbf{0.075}     & \textbf{0.195}      & \textbf{0.096} & \textbf{0.078}     \\ \hline
\multirow{3}{*}{Email-Eu*}  & \textit{TASBM} & 0.918     & 0.925      & 0.408 & 0.019     \\
  & \textit{STM}   & 0.722     & 0.732      & 0.242 & 0.480     \\
 & \textit{TagGen}   & 0.463     & 0.463      & 0.736 & 1.000     \\
  & \textit{MTM}   & \textbf{0.113}     & \textbf{0.080}     & \textbf{0.121} & \textbf{0.011}     \\ \hline
\multirow{3}{*}{SMS-A}      & \textit{TASBM} & 0.687     & 0.631      & \textbf{0.123} & 0.929     \\
      & \textit{STM}   & 0.668     & 0.653      & 0.362 & 0.067     \\
     & \textit{MTM}   & \textbf{0.096}     & \textbf{0.225}      & 0.168 & \textbf{0.003}     \\ \hline
\multirow{3}{*}{FBWall}     & \textit{TASBM} & 0.231     & 0.202      & 0.445 & 1.000     \\
     & \textit{STM}   & 0.431     & 0.429      & 0.513 & 1.000     \\
     & \textit{MTM}   & \textbf{0.030}     & \textbf{0.054}      & \textbf{0.028} & \textbf{0.008}  \\ \hline
\end{tabular}
\label{tab:KS}
\end{table}

We also examine the performance of certain distributions using the Kolmogorov-Smirnov (KS) test. In particular, we consider the in-degree, out-degree, inter-event time (IET), and timestamp distributions, and calculate the two sample KS test between the distributions in the original data and in the synthetic networks.
\cref{tab:KS} shows the average KS statistics (lower is better). We observe that \textit{MTM} consistently outperforms the baseline models for all distributions over all datasets, except the IET of SMS-A.

\subsection{Preserving Local Temporal Motif Statistics}
In this part, we evaluate how well the \textit{MTM} preserves the temporal motif statistics in the real-world networks compare to the baselines. 
For each dataset, we generate 10 synthetic networks using \textit{MTM} (with $l_{\mathrm{max}}=4$ and $\delta=1$ hour) and the baseline models. 
We exclude \textit{TagGen} model here because it generates output graphs as a sequence of static snapshots which does not contain temporal motif structures. 
We consider the inter-event time constraint $\Delta_C$ for counting temporal motifs, which requires that the time difference between each pair of consecutive events in the motif is less than $\Delta_C$.
Note that using a larger $\Delta_C$ threshold allows to discover more temporal motifs. However, a large
$\Delta_C$ has less power to control the relevance between consecutive events in the motif, and increases the computation cost. Given that the mean inter-event time in all datasets is less than 600 seconds (see \cref{tab:data}), we set $\Delta_C = 1$ hour and compute the counts of all 2-event, 3-event, and 4-event temporal motifs. We compare the number of motifs in the synthetic networks to the original input. 
We measure the difference by the mean squared relative error (MSRE),
$$
MSRE = \frac{1}{r} \sum^r_i \left(\frac{|\{M_{G_i}\}| - |\{M_{\mathrm{original}}\}|}{|\{M_{G_i}\}|}\right)^2,
$$
where $|\{M_{G_i}\}|$ is the number of motif instances identified in the $i$-th generated synthetic network, $|\{M_{\mathrm{original}}\}|$ is the number of motif instances in the original network, and $r=10$ is the number of synthetic networks generated.

\subsubsection{Total temporal motif counts}
\cref{tab:MSRE} shows the MSRE results based on the total number of 2-event, 3-event, and 4-event motifs.
\textit{MTM} significantly outperforms the baseline models across all the datasets.
\textit{TASBM} and \textit{STM} cannot preserve the 4-event motifs in \texttt{CollegeMsg} and \texttt{FBWall}.
\textit{MTM} generates synthetic networks with very similar motif counts where the MSRE is less than 0.5.
Note that \textit{TASBM} splits the input data into different time windows and generate synthetic network for each slice independently. Therefore, it achieves better MSRE for large datasets such as \texttt{FBWall} when compared to \textit{STM}. 
The MSRE of 3-event motifs is usually 10 times larger than the 2-event motifs and the error is even larger for the 4-event motifs. Larger size motifs are more complex and difficult to reproduce through generative models.

\begin{table}[!t]
\captionsetup{justification=centering}
\caption{\small The MSRE results for temporal motif counts.
}
\vspace{-2ex}
\small
\setlength\tabcolsep{1.5pt}
\centering
\begin{tabular}{|l|l|r|r|r|}
\hline
Data       & Model & 2-event  & 3-event   & 4-event   \\ \hline
\multirow{3}{*}{CollegeMsg} & \textit{TASBM} & 41.916   & 1214.121  & 17086.425 \\
 & \textit{STM}   & 296.274  & 3268.675  & 7704.935  \\
 & \textit{MTM}   & \textbf{0.004}    & \textbf{0.001}     & \textbf{0.035}     \\ \hline
\multirow{3}{*}{Email-Eu*}  & \textit{TASBM} & 0.634    & 2.885     & 5.852     \\
  & \textit{STM}   & 2.457    & 6.333     & 6.046     \\
  & \textit{MTM}   & \textbf{0.070}    & \textbf{0.152}     & \textbf{0.230}     \\ \hline
\multirow{3}{*}{SMS-A}      & \textit{TASBM} & 8.946    & 58.271    & 152.080   \\
     & \textit{STM}   & 515.802  & 3655.160  & 4420.164  \\
     & \textit{MTM}   & \textbf{0.0001}    & \textbf{0.010}     & \textbf{0.131}  \\ \hline
\multirow{3}{*}{FBWall}     & \textit{TASBM} & 3.605    & 34.467    & 171.514   \\
     & \textit{STM}   & 1569.383 & 10691.863 & 11694.559 \\
     & \textit{MTM}   & \textbf{0.0001}    & \textbf{0.021}     & \textbf{0.467}     \\ \hline
\end{tabular}
\label{tab:MSRE}
\vspace{-2ex}
\end{table}

\begin{figure}[!t]
\centering
\includegraphics[width=\linewidth]{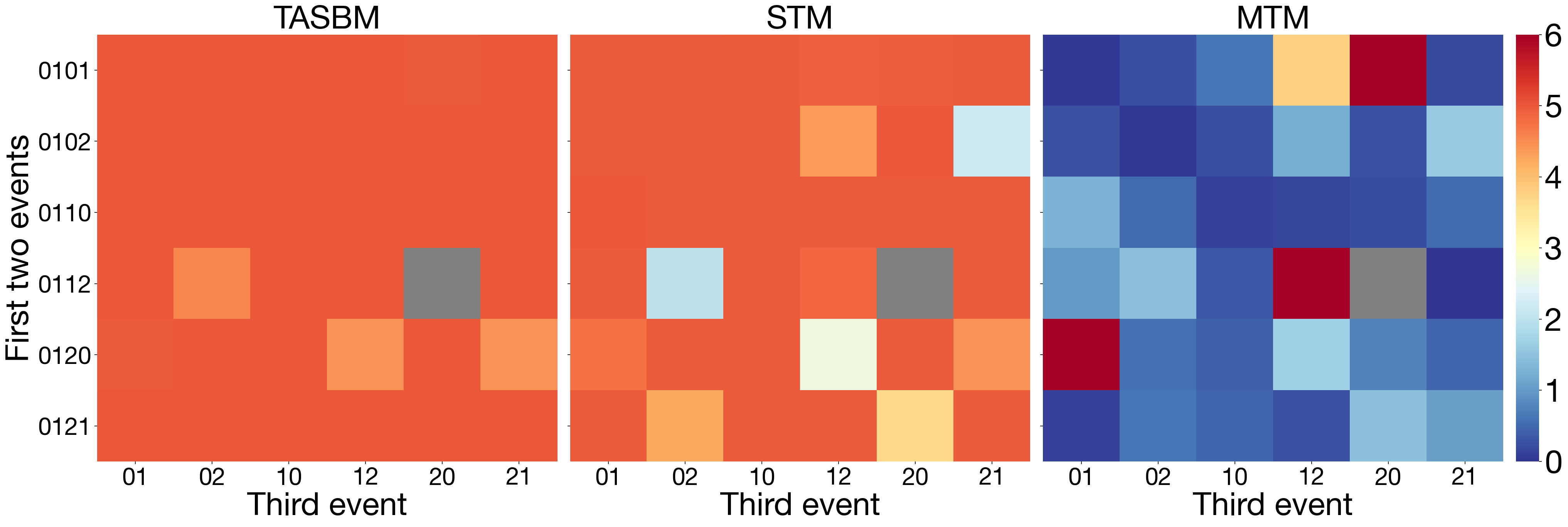}
\vspace{-2ex}
\caption{\small The MSRE results (color-coded) for all 3-node 3-event motifs in \texttt{CollegeMsg} network. Each block represents a type of motif; the row shows the notation of the first two events and the column shows the third event. Note that there are no \texttt{011220} motifs in the original network thus it does not have a valid MSRE value.}
\label{fig:heatmap}
\vspace{-3ex}
\end{figure}

\subsubsection{Motif spectrum counts}
Next, we dive into the motif spectrums and measure the MSRE for each type of motifs.
\cref{fig:heatmap} gives the MSRE results for all 3-node 3-event motifs in the synthetic \texttt{CollegeMsg} networks. Overall, our model gives 5 to 10 times smaller MSRE results than the baseline models. 
\textit{TASBM} assigns nodes to different activity groups and then generates the events between groups with different arrival rates. Since it does not utilize temporal motifs for graph generation, the MSRE are high for all types of motifs.
\textit{STM} selects the triangle motifs (\texttt{011202}, \texttt{012012}, \texttt{010221}, etc.) as a part of the atomic motif patterns, hence it yields smaller MSRE for those.
However, it cannot reproduce the distribution of other types of temporal motifs, thus yields high MSRE for 2-event, 3-event, and 4-event motifs in total.

\begin{figure}[!t]
\begin{subfigure}{0.49\linewidth}
\includegraphics[width=\linewidth]{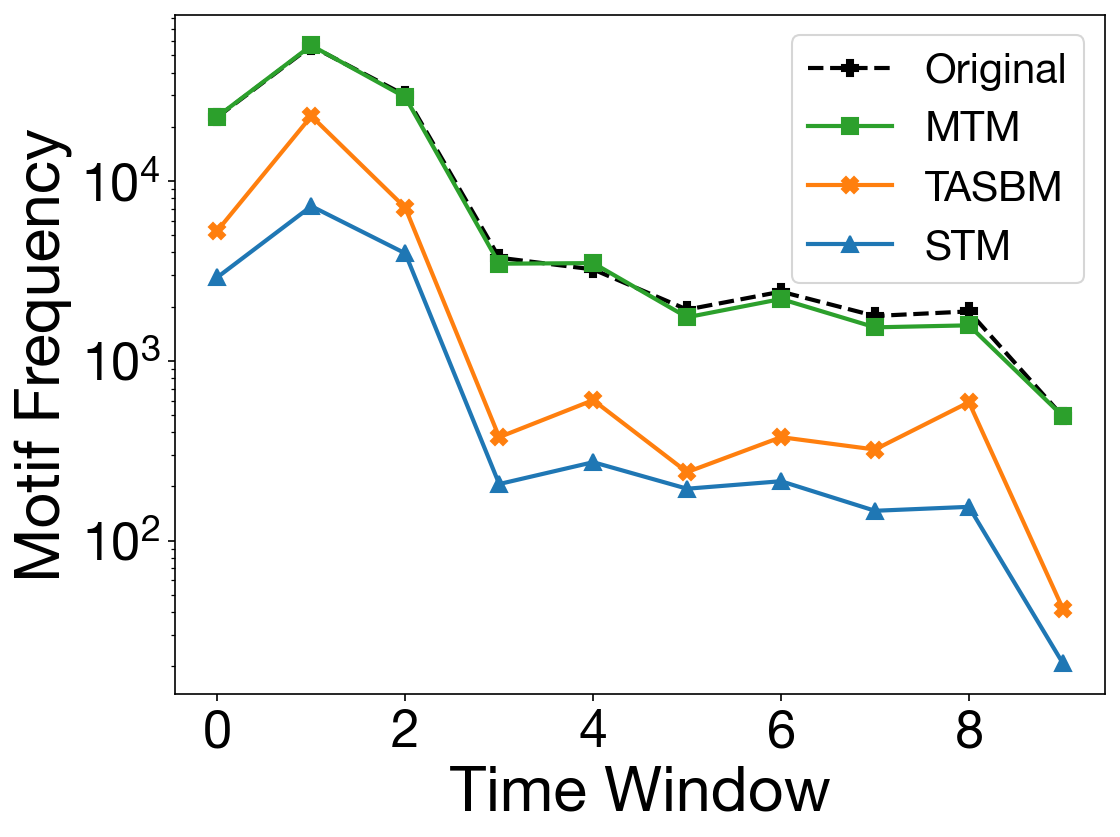}
\caption{\texttt{CollegeMsg}}
\end{subfigure}
\begin{subfigure}{0.49\linewidth}
\includegraphics[width=\linewidth]{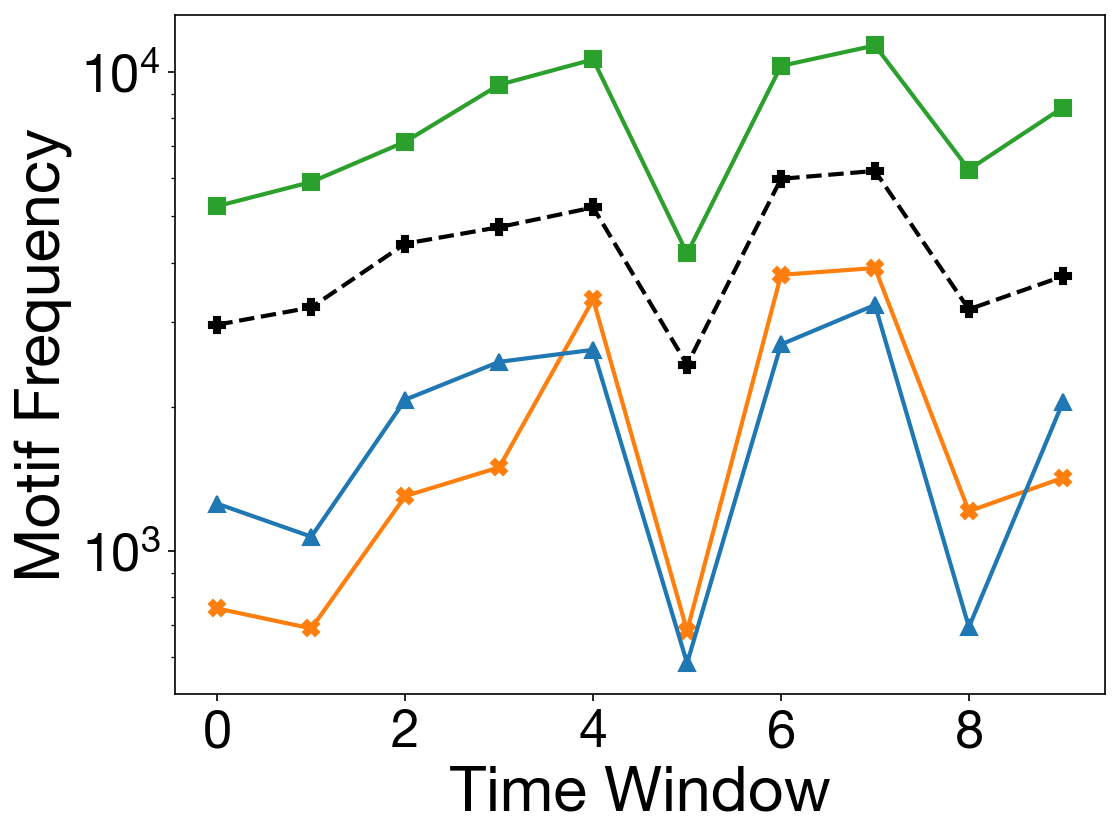}
\caption{\texttt{Email-EU*}}
\end{subfigure}
\vspace{-2ex}
\caption{\small The trend of the total number of temporal motifs over time in the original input and the synthetic networks.}
\label{fig:trend}
\end{figure}

\subsubsection{Distribution of motif counts over time}
Lastly, we investigate the distributions of the temporal motifs in the synthetic and real-world networks over time. 
We split each dataset into 10 equal size time windows, and generate synthetic networks using the events in each interval. We calculate the total number of 2-event, 3-event, and 4-event motifs in each time window and show the trends in~\cref{fig:trend} and ~\cref{fig:trend-appendix} of Appendix D.
Our model accurately simulates the trends in temporal motif counts over the time. The baseline models capture the trends to some extent but do not fit well to the actual numbers of temporal motifs, especially for the time windows with less counts of temporal motifs.
For example, the last time window of the \texttt{CollegeMsg} network has only 548 events (1\% of the entire data), but still contains a significant number of temporal motifs. The generative power of the baseline models degrades significantly in this window, especially for the \textit{STM} which does not generate any 3-event and 4-event motifs. Our model is robust to the changes in event density over time. 

\subsection{Runtime Analysis}

Here we examine the runtime performance of the \textit{MTM}. 
We first compare our model with baselines on all datasets. 
Note that we use a GPU to run \textit{TagGen} model.
\cref{tab:runtime} shows the average runtime of the 10  experiments.
We also give the runtime for motif counting to show how a hypothetical model that counts motifs would compare.
Our model takes significantly less time when compared to the baselines. The runtime of both baseline models increases drastically as the size of the input data increases. For larger datasets such as the \texttt{FBWall}, our model is 391 and 84 times faster than the \textit{TASBM} and \textit{STM}, respectively. 
In addition, our model is up to 231 times faster than motif counting computation.
This shows one of the key benefits of our model: we do not explicitly count the temporal motifs, instead we only compute the motif transition probabilities.

\begin{table}[!b]
\captionsetup{justification=centering}
\caption{\small The average runtime (seconds) for the generative models. We also show the number of events ($|E|$) for each dataset.}
\vspace{-2ex}
\small
\setlength\tabcolsep{1.5pt}
\centering
\begin{tabular}{|l|r|r|r|r|}
\hline
Data   & \texttt{Email-EU*}       & \texttt{CollegeMsg}   & \texttt{SMS-A}     & \texttt{FBWall}          \\ \hline
$|E|$ & 43035 & 59835 & 548182 & 876933 \\ \hline
\textit{TASBM}  & 24.7    & 36.8    & 11454.7  & 22100.9  \\
\textit{STM}   & 146.6   & 870.2   & 2565.1   & 4745.9   \\
\textit{TagGen} & 2147.4 & 15919.8 & N/A & N/A \\
Motif Counting    & 163.2     & 294.7     & 37957.7    & 1565.7    \\ 
\textit{MTM}    & \textbf{2.5}     & \textbf{2.8}     & \textbf{163.9}    & \textbf{56.5}    \\  \hline
\end{tabular}
\label{tab:runtime}
\vspace{-2ex}
\end{table}

\begin{figure}[!t]
\begin{subfigure}{0.49\linewidth}
\includegraphics[width=\linewidth]{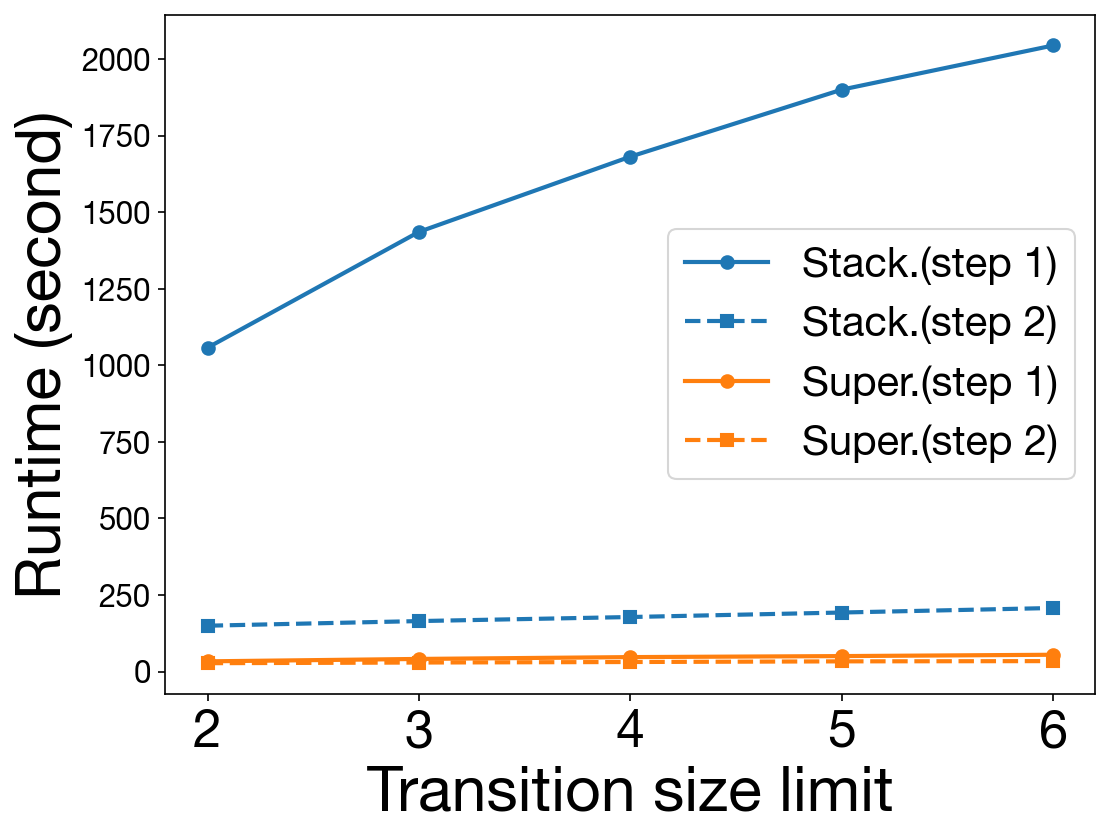}
\end{subfigure}
\begin{subfigure}{0.49\linewidth}
\includegraphics[width=\linewidth]{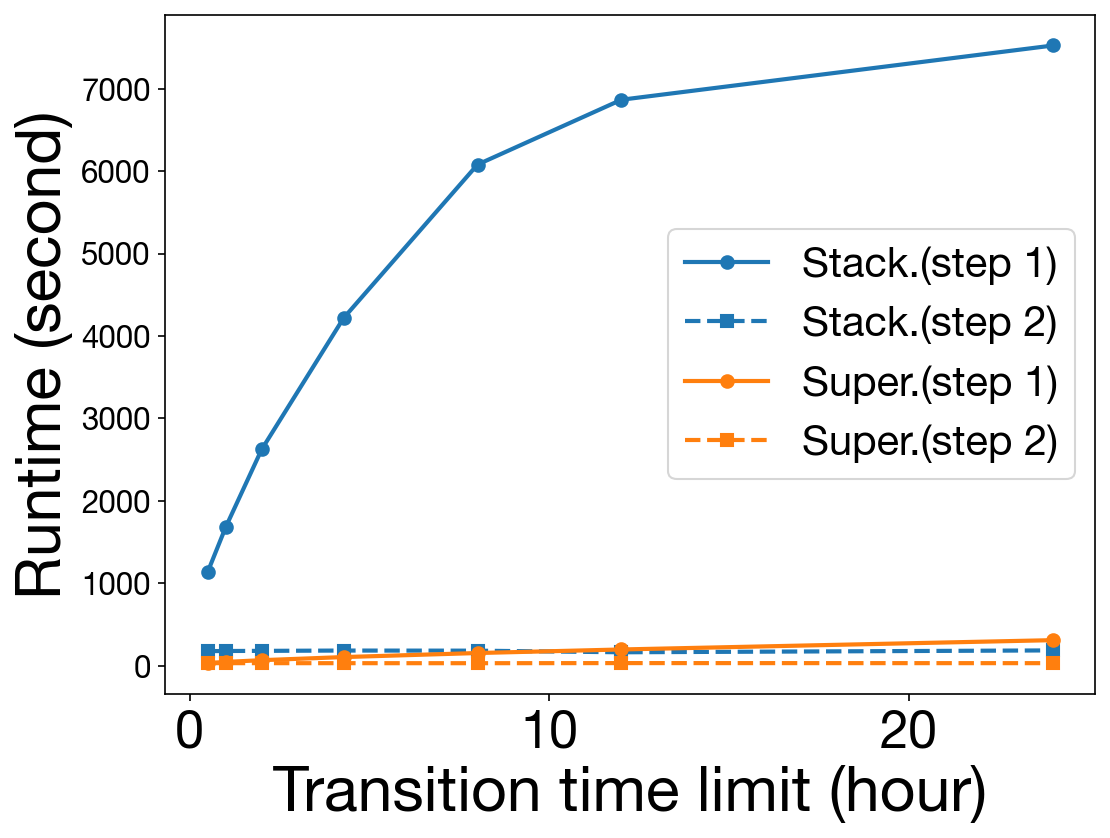}
\end{subfigure}
\vspace{-2ex}
\caption{\small The average runtime of  \textit{MTM} with respect to the transition size limit $l_{\mathrm{max}}$ (left) and the transition time limit $\delta$ (right).}
\label{fig:runtime_parameter}
\vspace{-4ex}
\end{figure}

We also evaluate the runtime of \textit{MTM} with respect to the two model parameters, $l_{\mathrm{max}}$ and $\delta$. We select two large datasets (\texttt{Super} {\tt User} and \texttt{StackOverflow}) with more than one million events, and measure the average runtime for identifying motif transition properties (step 1) and simulating motif transitions (step 2), respectively. To explore the impact of $l_{\mathrm{max}}$, we set $\delta=1$ hour and change $l_{\mathrm{max}}$ from 2 to 6.
Similarly, we set $l_{\mathrm{max}} = 4$ and calculate the average runtime with $\delta = [0.5, 1, 2, 4, 8, 12, 24]$ hours.
\cref{fig:runtime_parameter} shows the results. We observe that the runtimes increase gradually as the two parameters increase. However, the increase is less steep when $l_{\mathrm{max}} > 4$ and $\delta > 8$. One possible reason is that the higher-order temporal motifs are less common in real world networks, hence the additional cost of identifying motif transitions becomes less as the transition limits increase. Another reason is that larger $l_{\mathrm{max}} $ and $\delta$ values allow more events to be added to the transition processes. As a result, the number of cold events decreases, so as the number of active transition processes at each time step. This reduces the cost of identifying temporal motif transitions.

\section{Conclusion}
In this paper, we propose the \textit{MTM} model to generate a temporal network that preserves the global and local features of an input graph.
Our model first calculates five motif transition properties from the input graph, and then generates the synthetic networks by simulating the motif transition processes.
We evaluate the performance of our model on seven datasets from various domains. The experimental results show that our model is able to preserve the structural and temporal characteristics of the input graph, such as the size of the network, the average in- and out-degree, and the mean inter-event time. 
Our model shows superior performance on reproducing the temporal motif spectrums, even for the higher-order temporal motifs in large networks.
Last, but not least, our model 117 times faster (on average) than the baselines.

One potential opportunity for future research is developing the real-world applications of the motif transition models, such as generating temporal network with specified motif spectrums or detecting structural and temporal anomalies. 
Another possible direction is to consider other distributions to model the motif transition times, such as the temporal hawk process~\cite{hawkes1974cluster} and self-exciting cascading Poisson process~\cite{malmgren2008poissonian}.
It would also be interesting to explore other graph characteristics, such as the time-respecting paths and community structures, in the generated synthetic graphs and compare against the input network.

\section*{Acknowledgments}
This research was supported by NSF awards OAC-2107089 and IIS-2236789, and used resources
from the Center for Computational Research at the University at Buffalo~\cite{CCR}.

\pagebreak
\bibliography{paper,afosr}
\bibliographystyle{acm}

\pagebreak
\appendix

\section{Notations}
\begin{table}[!h]
\captionsetup{justification=centering}
\caption{\small Notation of symbols}
\vspace{-2ex}
\setlength\tabcolsep{1.5pt}
\centering
\begin{tabular}{|l|l|}
\hline
Symbol     & Description \\ \hline \hline
$G = (V, E)$ & temporal graph \\ \hline 
& \\ [-1em]
$\overline{G} = (V, \overline{E})$ & the static projection of $G$ \\ \hline
& \\ [-1em]
$M^l_i$      & temporal motif with $l$ events                      \\ \hline
& \\ [-1em]
$\mathcal{T}(M^l_i {\rightarrow} M^{l+1}_j)$ & motif transition from $M^l_i$ to $M^{l+1}_j$ \\ \hline
& \\ [-1em]
$\mathcal{T}(M^1 {\rightarrow} \dots {\rightarrow} S)$ & motif transition process \\ \hline
& \\ [-1em]
$P(M^{l+1}_j | M^l_i)$ & transition probability \\ \hline
& \\ [-1em]
$\lambda(M^l_i {\rightarrow} M^{l+1}_i)$ & transition rate \\ \hline
$\Delta_t$ & transition time\\ \hline
$CE$ & set of cold events\\ \hline
$K_{CE}$ & degrees of cold events\\ \hline
$T_{CE}$ & timestamps of cold events\\ \hline
$\mu$ & average number of edges in \\ 
& motif transition processes\\ \hline
$l_{\max}$ & transition size limit \\ \hline
$\delta$ & transition time limit \\ \hline
$|\mathbb{T}|$ & total number of transition types \\ \hline
$|T|$ & the timespan of the input graph \\ \hline
\end{tabular}
\label{tab:notation}
\vspace{-3ex}
\end{table}

\section{Data Statistics}
\begin{table}[!h]
\captionsetup{justification=centering}
\caption{\small Statistics of the temporal network datasets. 
}
\small
\setlength\tabcolsep{1.5pt}
\centering
\begin{tabular}{|l|r|r|r|r|r|}
\hline
Name     & Nodes & Edges & Events & Timespan (days) & mean IET (sec.) \\ \hline
\texttt{CollegeMsg}      & 1.90K & 20.3K & 59.8K  & 193 & 273.1                     \\
\texttt{Email-EU}               & 986   & 24.9K   & 332K & 803    & 209.0                     \\
\texttt{Email-EU*}              & 80   & 1184 & 43.0K   & 500 & 514.2                    \\
\texttt{SMS-A}          & 44.4K & 69.0K    & 548K   & 338 & 53.3                      \\ 
\texttt{FBWall}             & 47.0K & 274K  & 877K   & 1560 & 152.9                     \\
\texttt{SuperUser}          & 194K & 925K   & 1.44M  & 2773 & 166.0                      \\ 
\texttt{StackOverflow}  & 260K & 4.15M & 6.35M & 886& 12.0 \\ \hline
\end{tabular}
\label{tab:data}
\end{table}

We use several real-world temporal networks from various domains.
\cref{tab:data} give the statistics.
In addition to the number of nodes, edges, and events, we give timespan and the mean inter-event time (i.e., average of the time intervals between all pairs of consecutive events) for each dataset. 
\texttt{CollegeMsg} \cite{snap} and \texttt{SMS-A} are phone message networks, in which an event $(u, v, t)$ represents a message sent from $u$ to $v$ at time $t$.
\texttt{Email-EU} is the emails between members of a European research institution \cite{snap}, in which an event $(u, v, t)$ indicates an email sent from person $u$ to person $v$ at time $t$
(we also include the \texttt{Email-EU*} network used by \cite{porter2022analytical}, which contains 80 densely connected nodes in largest connected component of the original \texttt{Email-EU} data).
\texttt{FBWall} contains the posts between users on Facebook in the New Orleans region~\cite{viswanath2009}, where an event $(u, v, t)$ denotes user $u$ posted on the user $v$'s wall at time $t$.
\texttt{SuperUser} and \texttt{StackOverflow} are the interaction networks from stack exchange websites~\cite{snap}, where an event $(u, v, t)$ stands for user $u$ posted an answer/comment on user $v$'s question/answer at time $t$.
The time resolution of all the networks is one second.

\section{All 3-event Motifs}
\begin{figure}[!h]
\vspace{-2ex}
\centering
\includegraphics[width=0.98\linewidth]{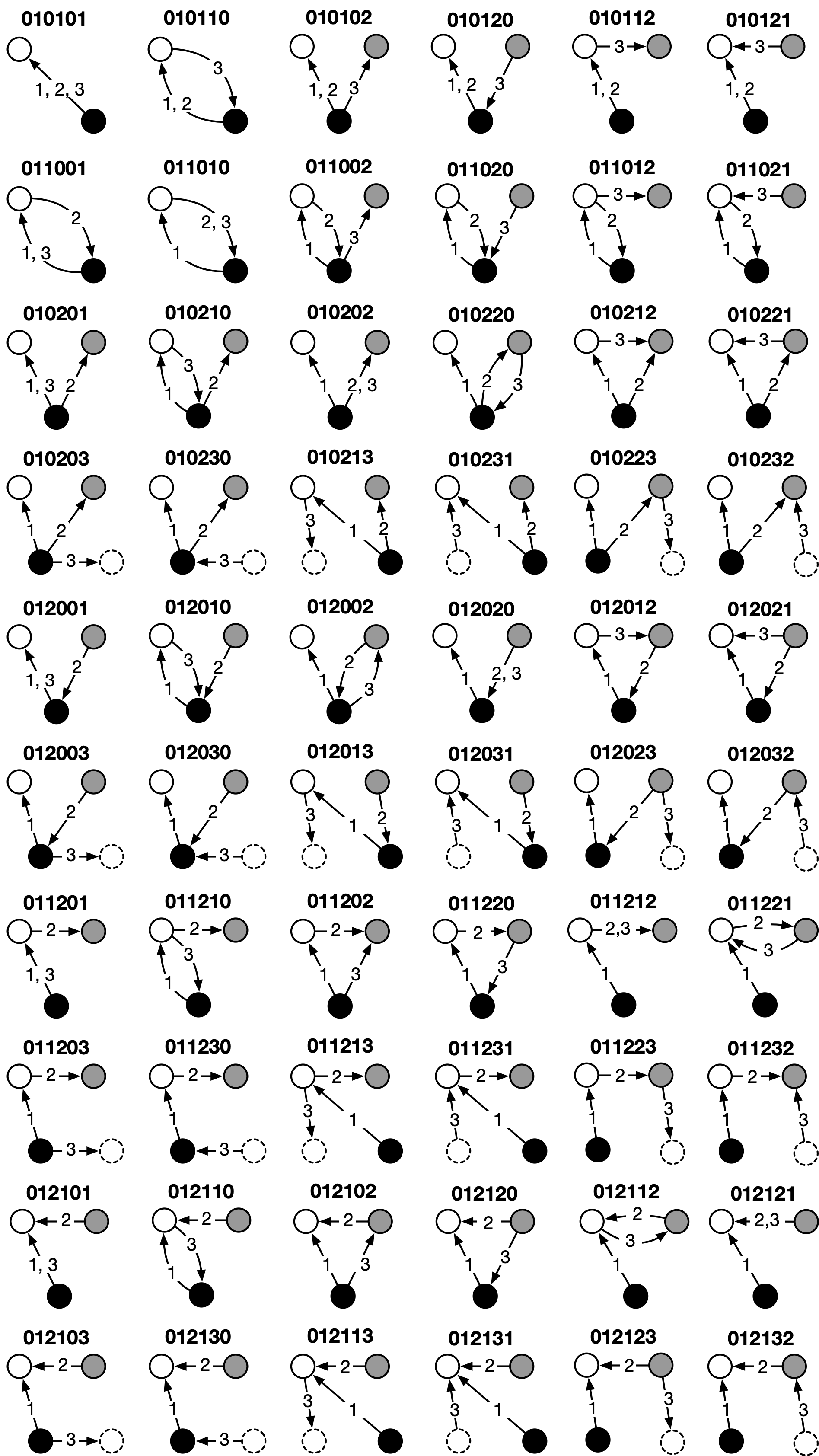}
\caption{\small All 60 types of 3-event motifs.}
\label{fig:m60}
\vspace{-3ex}
\end{figure}

Here we list all 60 types of 3-event temporal motifs in~\cref{fig:m60}.

\section{Additional Results}
\begin{figure}[!b]
\vspace{-4ex}
\begin{subfigure}{0.49\linewidth}
\includegraphics[width=\linewidth]{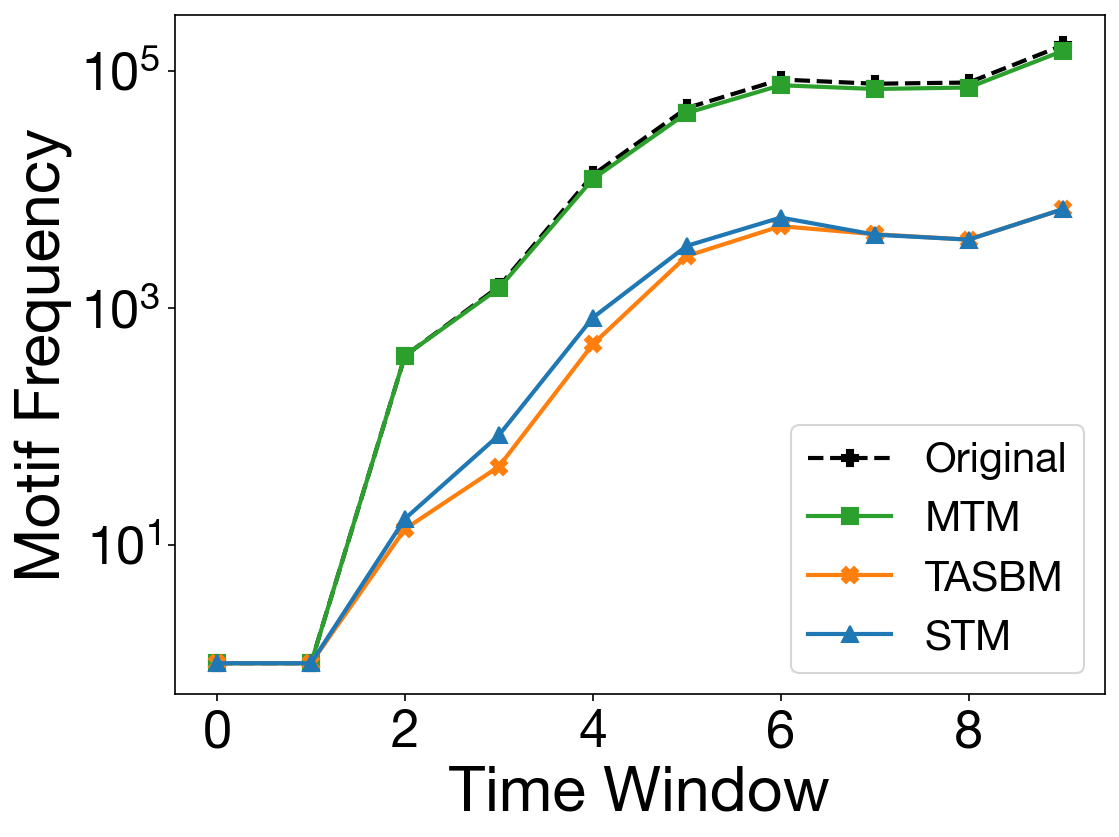}
\caption{\texttt{FBWall}}
\end{subfigure}
\begin{subfigure}{0.49\linewidth}
\includegraphics[width=\linewidth]{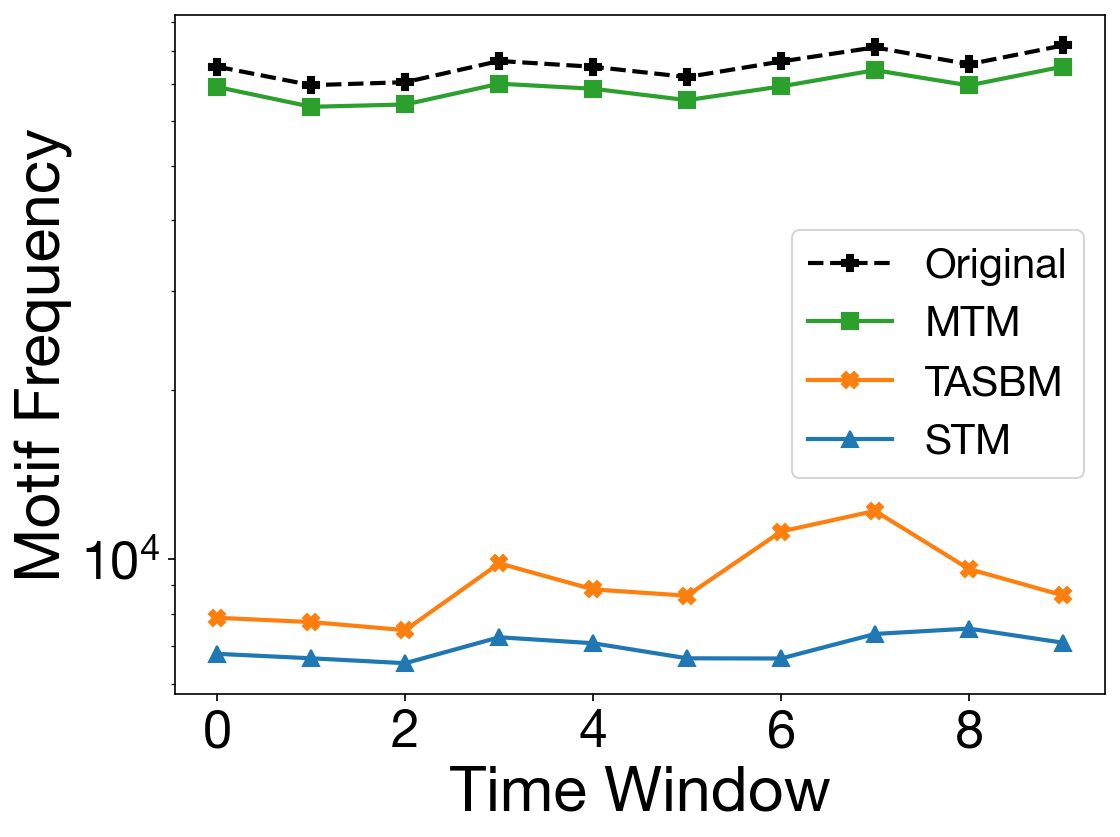}
\caption{\texttt{SMS-A}}
\end{subfigure}
\vspace{-2ex}
\caption{\small The trend of the total number of temporal motifs over time in the original input and the synthetic networks.}
\label{fig:trend-appendix}
\end{figure}

\end{document}